\documentclass[fleqn,usenatbib]{mnras}

\usepackage{newtxtext,newtxmath}
\usepackage[T1]{fontenc}

\DeclareRobustCommand{\VAN}[3]{#2}
\let\VANthebibliography\thebibliography
\def\thebibliography{\DeclareRobustCommand{\VAN}[3]{##3}\VANthebibliography}

\usepackage{graphicx}	% Including figure files
\usepackage{amsmath}	% Advanced maths commands
\usepackage{bm}
\title[circumbinary disc accretion]{Circumbinary accretion as a diagnostic for binary--disc misalignment}

\author[Smallwood et al.]{Jeremy L. Smallwood,$^{1,2}$\thanks{E-mail: smallj2@ou.edu}
Ya-Ping Li,$^{3}$
Hongping Deng,$^{3}$
and
Alessia Franchini$^{4}$
\\
 $^{1}$Institute of Astronomy and Astrophysics, Academia Sinica, Taipei 10617, R.O.C.\\
 $^{2}$Homer L. Dodge Department of Physics and Astronomy, The University of Oklahoma, Norman, OK 73019, USA\\
 $^{3}$Shanghai Astronomical Observatory, Chinese Academy of Sciences, Shanghai 200030, People's Republic of China\\
 $^{4}$Institut f\"{u}r Astrophysik, Universit\"{a}t Z\"{u}rich, Winterthurerstrasse 190, 8057 Z\"{u}rich, Switzerland\\
}

\date{Accepted XXX. Received YYY; in original form ZZZ}

\pubyear{2024}

\begin{document}
\label{firstpage}
\pagerange{\pageref{firstpage}--\pageref{lastpage}}
\maketitle

% Abstract of the paper
\begin{abstract}
 Binary star systems can accrete material originating from a circumbinary disc. Since it is common for the circumbinary disc to be tilted with respect to the binary orbital plane, we test whether the accretion dynamics can be a diagnostic for binary-disc misalignment. We present hydrodynamical simulations to model the accretion flow from a circumbinary disc around an eccentric binary with initial tilts ranging from $0^\circ$ to $180^\circ$ in increments of $15^\circ$. Based on the initial tilt, the circumbinary disc will align towards three different configurations: prograde coplanar, polar, or retrograde coplanar. For discs with initial tilts evolving towards prograde coplanar alignment, the accretion rates onto the primary and secondary stars exhibit alternating preferential accretion. Circumbinary discs evolving towards polar alignment exhibit no alternating preferential accretion onto the binary unless the initial tilt is close to the critical tilt that sets the boundary between coplanar or polar alignment. Such cases cause strong disc warping, leading to disc breaking. The inner disc becomes eccentric, leading to alternating preferential accretion onto the binary. As the break propagates outward, the disc tilt damps towards a polar state and the disc eccentricity decreases. As the disc re-circularizes, the accretion rate transitions back from alternating preferential accretion to non-alternating accretion. Lastly, no alternating preferential accretion exists for discs undergoing retrograde coplanar alignment. From the summary of the accretion rates from our suite of SPH simulations, it is evident that the accretion rate evolution can be affected by the initial tilt and subsequent evolution of the circumbinary disc.
\end{abstract}

% Select between one and six entries from the list of approved keywords.
% Don't make up new ones.

%%%%%%%%%%%%%%%%%%%%%%%%%%%%%%%%%%%%%%%%%%%%%%%%%%

%%%%%%%%%%%%%%%%% BODY OF PAPER %%%%%%%%%%%%%%%%%%

\section{Introduction}

The orbital evolution of a binary system can be strongly influenced by the interaction between the binary components and the surrounding gas \citep{Bate2002}. The gas settles in a circumbinary disc  and the binary torque carves a wide cavity in the disc, possibly limiting the process of accretion \citep{Artymowicz1994}. \cite{Artymowicz1996} demonstrated, using 3D hydrodynamical simulations that the gas can leak into the cavity through high-speed streams originating along the outside of the cavity and spiraling inward to the central binary. The dynamics of these streams and the subsequent formation of circumstellar discs has been investigated in detail using more modern simulations \citep{Shi2012,DOrazio2013,Miranda2017}. 

The binary can exchange angular momentum and energy with the surrounding disc via accretion and gravitational torques \citep{Farris2014,Roedig2012} and this translates into an evolution of its orbital parameters: eccentricity and semi-major axis. The evolution of both parameters has been investigated recently by a number of works, using both 2D fixed binary orbit simulations \citep{Munoz2019,Munoz2020b,Duffell2020,Tiede2020} and 3D live binary simulations with isothermal discs \citep{Ragusa2016,Heath2020,Franchini2022,Franchini2023} and massive self-gravitating discs \citep{Roedig2012,Franchini2021,Franchini2024}. The vast majority of these simulations are scale free so their results can in principle be applied to both the massive black hole binary regime and the binary star case. We here consider the latter.

%The long-term evolution of a binary star system is dictated by the interaction between the two central bodies and the surrounding gas via accretion and gravitational torques \citep{Artymowicz1983,Bate2002,Farris2014,Gerosa2015,Duffell2015,Munoz2020b}. The tidal torque of the binary exerted on the disc can facilitate, and perhaps limit, the process of accretion. Usually, the accretion occurs in short, high-speed streams originating along the outside of the cavity and spiraling inward to the central binary. 

% \af{I think the following 4 paragraph below need to be re-arranged. I would divide between observations and theory and then between unequal-mass and equal mass systems.}

The mass accretion rate onto each star, $\dot{M}_{\rm acc}$, is instrumental for the study of the accretion disc evolution.
Observationally, $\dot{M}_{\rm acc}$ can be estimated by measuring the flux of continuum and line emissions resulting from the shock of infalling gas from a disc onto the central star along the stellar magnetic field lines \cite[e.g.,][]{Calvet1998}. Observations reveal accretion rates onto binary star systems are comparable to that of accretion rates onto single T Tauri stars \citep{White2001}, which are typically in the range of $\sim 10^{-10} - 10^{-7}\, \rm M_{\odot}/yr$ \citep{Valenti1993,Gullbring1998,Calvet2004,Ingleby2013}. Furthermore, observations of binary star systems suggest that the more massive star typically exhibits a higher accretion rate compared to the less massive one \citep{White2001}. Additionally, \cite{Manara2012} noted an increase in $\dot{M}_{\rm acc}$ with stellar mass, which decreases over evolutionary time, based on observations of the Orion Nebula Cluster. 

Variations in the circumbinary gas disc and gas streams structure depend significantly on binary eccentricity and mass ratio \citep{Siwek2022}. High eccentricity binaries with mass ratios close to unity exhibit notable accretion rate modulations, occurring on timescales similar to the binary orbital period ($P_{\text{orb}}$), driven by time-varying gravitational forces. Additionally, these binaries lead to substantial eccentricity in the circumbinary disc eventually dissipating and reaching a steady value of $\sim 0.3$ \cite[e.g.,][]{Lubow2000a,Papaloizouetal2001,MacFadyen2008,Cuadra2009,DOrazio2013,Miranda2017,Munoz2016,Thun2017,Munoz2019,Munoz2020b,Duffell2024}. Therefore, the eccentric circumbinary disc's apsidal precession around such binaries results in preferential alternating accretion of material between the binary members over time \citep{Munoz2016}.  On the contrary, circular equal-mass binaries do typically exhibit accretion rate modulations on both the binary orbital period and on the timescale related to the disc inner edge orbital motion, i.e. $5\, \rm P_{orb}$ \citep{Franchini2023,Lai2023}.
%{\bf On the contrary, circular binaries with nearly equal-mass ratios will exhibit accretion rate modulations occurring on timescales of $5\, \rm P_{orb}$ \citep{Lai2023}. }

The preferential alternating accretion phenomenon can explain why such observations, like the eccentric T Tauri  binary TWA 3A, shows a dominant accretion onto the primary \citep{Tofflemire2019}, while most simulations suggest that the accretion onto secondary is preferred \citep{Bate2000,Farris2014,Munoz2020a}. 
% Eccentric circumbinary accretion discs around binary massive black holes have been found through two-dimensional grid-based hydrodynamics simulations \citep{MacFadyen2008}. 
\cite{Dunhill2015} modelled a coplanar circumbinary disc around the eccentric binary HD 104237 and found the accretion rate on to the binary results in a periodic accretion variability. Furthermore, \cite{Ragusa2020} found similar eccentric behaviour in coplanar circumbinary discs around a circular binary utilizing 3-dimensional hydrodynamical simulations.

Early hydrodynamics simulations conducted by \cite{Bate1997} and \cite{Bate2000} revealed that the secondary star can manifest a significantly higher accretion rate than the primary under conditions where the gas possesses substantial angular momentum. This phenomenon is thought to occur because the secondary star occupies a larger orbit and thus orbits closer to the inner edge of the disc, facilitating greater gas accretion. Nevertheless, if this were universally true, the binary mass ratio would trend towards unity \citep{Hanawa2010}, yet observations indicate the prevalence of unequal-mass binaries \citep{Duquennoy1991}. Contradictorily, grid-based simulations by \cite{Ochi2005} indicated a higher accretion rate for the primary star. Furthermore, \cite{Hanawa2010} computed gas accretion from a circumbinary disc utilizing a more precise scheme to resolve the centrifugal balance of a gas disc against gravity. Their findings suggested that the mass accretion rate typically favors the primary star.
Preferential accretion onto one of the binary component has been the subject of more recent studies \citep{Duffell2020,Siwek2022,Franchini2024} that presented similar results in terms of its dependence on the initial binary mass ratio. In particular, these studies confirm the preferential accretion onto the secondary star as long as the binary eccentricity is $<0.8$ and the mass ratio is $<0.3$.  In their study, \cite{Ceppi2022} performed smoothed-particle hydrodynamical simulations of hierarchical triple star systems, focusing on the accretion rates within a coplanar circumtriple disc. Their results indicate that if the inner binary's mass exceeds that of the tertiary component, the typical differential accretion process, in which the secondary star accretes more material, is impeded. However, when the inner binary is less massive than the third body, this differential accretion scenario is enhanced.

%The extensively studied dynamics of coplanar circumbinary disc systems involve the initial formation of a central cavity within the circumbinary disc, induced by the binary system's tidal torques \citep{Artymowicz1994, Miranda2015}. Subsequent gas inflow into this cavity in the form of gas streams, observed in numerous studies, is crucial for circumstellar disc formation and evolution \citep[e.g.,][]{Artymowicz1996, Gunther2002, Nixon2012, Shi2012, DOrazio2013, Farris2014, Miranda2017, Munoz2019, Mosta2019}. 

There is little known about the accretion evolution onto binaries from misaligned circumbinary discs. Recently, \cite{Smallwood2022} analysed the accretion of material from a polar circumbinary. They found that the binary does not excite significant eccentricity in the disc, resulting in no alternating preferential accretion. This is because tidal torques due to Lindblad resonances on a polar circumbinary disc approach zero in the limit of unity binary eccentricity \citep{Lubow2018}. In this work, we use hydrodynamical simulations to model circumbinary discs with misalignments with respect to the binary orbital plane in the range $0^\circ-180^\circ$ in steps of $15^\circ$ and track the accretion evolution onto the binary. At a given radius, tidal torques are weaker for a misaligned circumbinary disc, giving rise to smaller central cavities \citep{Lubow2015,Miranda2015, Franchini2019b}, which may impact the accretion dynamics. We find that the accretion pattern may be used to infer the binary-disc misalignment. 

The structure of the paper is as follows. We detail the initial setup of the hydrodynamical simulation in Section~\ref{sec::setup}. We present the results in Section~\ref{sec::results} and give a discussion in Section~\ref{sec::discussion}. Finally, we draw our conclusions in Section~\ref{sec::conclusion}.

\section{Methods}
\label{sec::setup}

\subsection{Binary and disc setup}
To model the accretion of material from misaligned circumbinary discs, we use the smoothed particle hydrodynamical code {\sc Phantom} \citep{Price2018}. {\sc Phantom} has been successful in  modeling misaligned circumbinary discs \citep[e.g.,][]{Nixon2012,Nixon2013,Dougan2015,Facchini2018, Aly2020,Smallwood2020a,Smallwood2022}. The discs within our hydro model are in the bending wave regime, meaning that the disc aspect ratio, $H/r$, is larger than the \cite{Shakura1973} $\alpha$-viscosity parameter, which is appropriate for protoplanetary discs \cite[e.g.,][]{Hueso2005,Rafikov2016,Ansdell2018}. In this regime, warps induced in the disc by the binary torque propagate as  bending waves with a vertical averaged speed $c_{\rm s}/2$ \citep{Papaloizou1995}, where $c_{\rm s}$ is the sound speed.  

We set up an equal-mass eccentric binary with a semi-major axis $a = 1$ and initial eccentricity $e_{\rm b} = 0.5$. The binary separation and eccentricity are allowed to evolve in time. We employ a Cartesian coordinate system ($x$,$y$,$z$), where the $x$-axis aligns with the direction of the binary eccentricity vector, and the $z$-axis aligns with the direction of the binary angular momentum vector. We model the equal-mass binary as a pair of sink particles with a total mass $M_1 + M_2 = M$, where $M_1$ and $M_2$ are the masses of the primary and secondary binary components, respectively. The binary accretion radii are set to $r_{\rm acc,1} = r_{\rm acc,2} = 0.25a$. The sink accretion radius is considered a hard boundary, where the accreted particles' mass and angular momentum are added to the sink \citep{Bate1995}. \cite{Smallwood2022} performed a resolution study about how the accretion radii affect the accretion rate. They found that the accretion rate with the smallest accretion radius, $0.01a$ (comparable to the size of the star), was similar to the accretion rate for $r_{\rm acc} = 0.25a$. We therefore chose a larger sink radius in order to decrease the computational time.%A larger sink accretion radius will decrease the computational time.  

The circumbinary disc is modelled with initially $1\times 10^6$ equal-mass SPH particles with a total disc mass $M_{\rm d} = 10^{-3}\, \rm M$. The particles are radially distributed from the inner disc radius, $r_{\rm in} = 4a$, to the outer disc radius, $r_{\rm out} = 7a$. A coplanar disc around a binary would have an inner truncation radius of about $3a$ \citep{Artymowicz1994}, whereas a tilted circumbinary disc can radially extend much closer to the binary orbit due to the weaker binary gravitational torque experienced by the off-plane material. %binary torque being weaker at a given radius. 
The observations of the polar disc around HD 98800 BaBb reveal indeed an inner disc radius of $\sim 1.6a$ \citep{Franchini2019b}. The chosen value of $r_{\rm in}=4a$ exceeds the tidal truncation limit for both extremes: coplanar and polar configurations. This allows the material to initially viscously drift inward  down to the effective truncation radius before the disc settles into a quasi-steady state \cite[e.g.,][]{Smallwood2022}.
 A smaller initial disc inner edge would only cause an initial transient in the accretion rate onto the binary as the mass within the truncation radius is quickly accreted \cite[e.g.,][]{Smallwood2021b}. This transient would furthermore cause the resolution of the simulations to decrease as the disc loses mass. We find the long term accretion rate from a coplanar disc starting at $2a$ or at $4a$ to be the same. 
%Initiating the inner disc edge too close to the binary could lead to an initially exaggerated accretion rate onto the binary, leading to the loss of disc mass \cite[e.g.,][]{Smallwood2021b}. {\bf Once the disc reached a quasi-steady state, the accretion rate became consistent with the case of $r_{\rm in} = 4a$ simulations.} 
Additionally, the disc mass is not large enough for self-gravity to be important.

The gas surface density profile is initially a power-law distribution given by
 \begin{equation}
     \Sigma(r) = \Sigma_0 \bigg( \frac{r}{r_{\rm in}} \bigg)^{-p},
     \label{eq::sigma}
 \end{equation}
where $\Sigma_0 =  6.75\times10^{-10}\, \rm M/a^2$ is the density normalization (defined by the total mass), $p$ is the power law index, and $r$ is the spherical radius. We set $p=3/2$. We adopt the locally isothermal equation of state of \cite{Farris2014} and set the sound speed $c_s$ to be
\begin{equation}
    c_{\rm s} = c_{\rm s0}\bigg( \frac{a_{\rm b}}{M_1 + M_2} \bigg)^q \bigg( \frac{M_1}{R_1} + \frac{M_2}{R_2}\bigg)^q,
    \label{eq::EOS}
\end{equation}
where $R_{\rm 1}$ and  $R_{\rm 2}$ are the radial distances from the primary and secondary star, respectively, and $c_{\rm s0}$ is a constant with dimensions of velocity This approach to the sound speed distribution ensures that the temperatures around the circumprimary and circumsecondary discs are primarily influenced by the irradiation from the primary and secondary stars, respectively. For $R_1(R_2) \gg a$, the sound speed is set by the distance from the binary centre of mass. The disc thickness is scaled with radius as
\begin{equation}
    H = \frac{c_{\rm s}}{\Omega} \propto r^{3/2-q}, 
 \end{equation}
where $\Omega = \sqrt{GM/r^3}$ and $q = 3/4$.  We set an initial gas disc aspect ratio of $H/r = 0.1$ at $r = r_{\rm in}$. The \cite{Shakura1973} viscosity, $\alpha_{\rm SS}$, prescription is given by 
\begin{equation}
    \nu = \alpha_{\rm SS} c_{\rm s} H,
\end{equation}
where $\nu$ is the kinematic viscosity. In order to simulate
$\alpha_{\rm SS}$, we use the artificial viscosity $\alpha^{\rm av}$ prescription in \cite{Lodato2010} given as
\begin{equation}
\alpha_{\rm SS} \approx \frac{\alpha_{\rm AV}}{10}\frac{\langle h \rangle}{H}.
\end{equation}
 We take the \cite{Shakura1973} $\alpha_{\rm SS}$ parameter to be $0.01$. The circumbinary disc is initially resolved with average smoothing length per scale height $\langle h \rangle / H = 0.20$. With the above prescription,  $\langle h \rangle / H$ and $\alpha_{\rm SS}$ are constant over the radial extent of the disc \citep{Lodato2007}.  Note, that $\langle h \rangle / H$ can vary due to the formation of warps in the disc \citep{Fairbairn2021,Deng2022}.

 The material that flows into the cavity from the circumbinary disc, rather than being flung back to its inner edge, will ultimately be accreted onto the sinks. This occurs on a timescale linked to the viscous timescale within each circumstellar disc. However, this timescale remains largely uncertain, as there are currently no reliable estimates for the viscosity parameter in these types of discs. Our chosen equation-of-state described in Eq.~(\ref{eq::EOS}) allows for the formation of circumstellar discs around each component. However, the resolution of these circumstellar discs formed by the streams of material leaking from the circumbinary disc is quite low. Indeed in SPH simulation it is very challenging to resolve the dynamics in low density regions, such as the cavity carved by the binary into the disc. %the accretion of circumbinary material is quite low since we use 3D SPH. 
Since the two circumstellar disc have low resolution, their viscous timescale is artificially short, leading to rapid accretion of the circumstellar disc onto the binary component. In order to resolve %simulate 
the circumstellar discs in {\sc phantom} %supplied by circumbinary gas on longer timescales, 
we would need to employ the method outlined in \cite{Smallwood2021a}, where they artificially decrease the sound speed near each binary component. This increases the viscous timescale in the circumstellar discs, enabling material accumulation around the binary components. However, employing this method would prohibit us from simulating over the extended timescales necessary to model the quasi-steady state accretion rate from the circumbinary disc to the binary as the resolution of bound material on small-scales is computationally intensive. Hence, we decide to not resolve %opt not to depict 
the formation of such discs in our hydrodynamical simulations.%, as the increased number of particles required to resolve circumstellar disc formation around each binary component would substantially prolong the computational time. 
The buffering effect of circumstellar discs on accretion onto the stars is not considered, as the circumstellar discs are not simulated. Moreover, the likelihood of forming long-lived circumstellar discs decreases with increasing eccentricity of the binary components and/or decreasing binary separation due to tidal truncation \citep[e.g.,][]{Artymowicz1994}.

We explore several disc misalignment angles, uniformly sampled between coplanar and retrograde in increments of $15^{\circ}$.

\begin{figure*} 
\centering
\includegraphics[width=1\columnwidth]{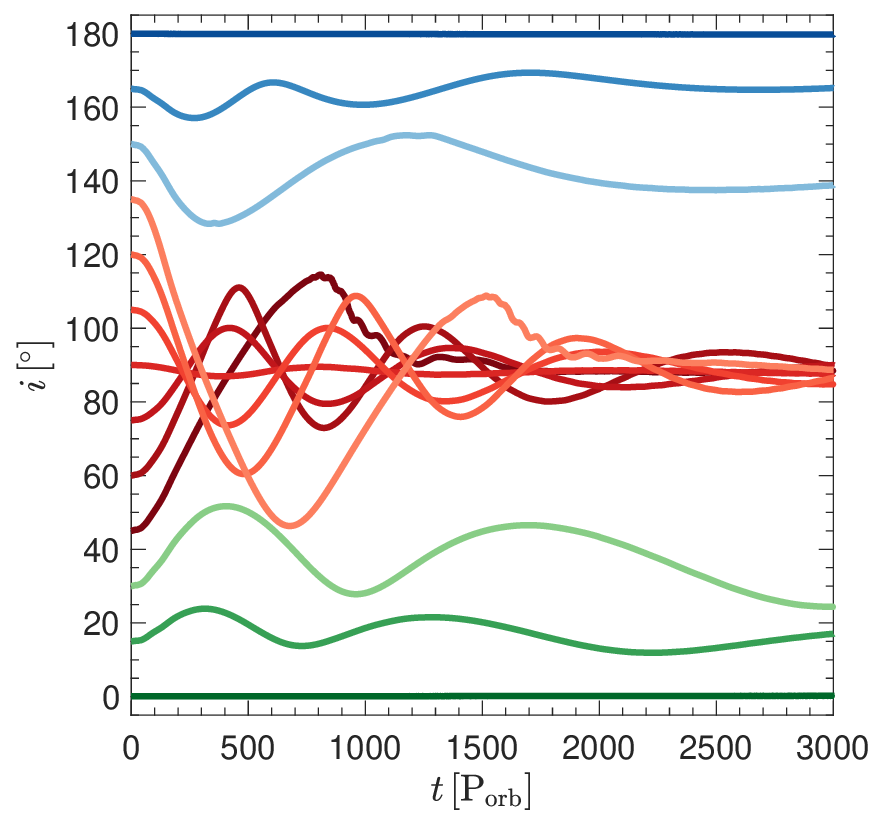}
\includegraphics[width=1\columnwidth]{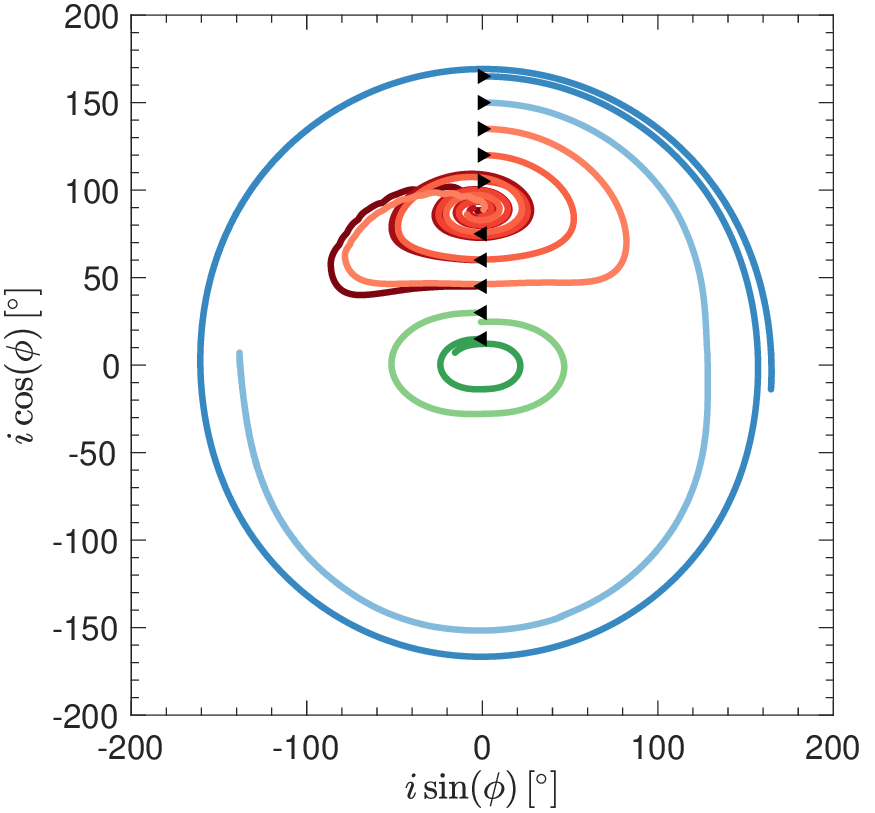}
\caption{Left panel: The evolution of the tilt, $i$, as a function of time in units of initial binary orbital period, $P_{\rm orb}$. The colours denote the initial tilt of the circumbinary disc with greens evolving coplanar prograde, reds evolving polar, and blues evolving coplanar retrograde. Right panel: disc evolution in the $i\cos \phi $--$i\sin \phi$ phase space. The black triangles represent $t = 0, \rm P_{orb}$, with each triangle pointing in the direction of evolution. Note that simulations with $i_0 = 0^\circ$, $90^\circ$, $180^\circ$ are not shown because the longitude of the ascending node is undefined for coplanar disc orientations, and there is very little precession when the disc is initially polar.}
\label{fig::tilt}
\end{figure*}

\subsection{Analysis}

The analysis of the simulations presented in this work is focused on the connection between the disc properties and the accretion rate onto the binary components. In particular, we want to understand whether the fact that the disc is circulating, therefore aligning/counter-aligning with the binary orbital, or librating, therefore undergoing polar alignment, affects the preferential accretion onto one of the binary components. 
%To analysis the SPH simulations, w
We operate within a framework defined by the instantaneous values of the binary eccentricity vector, $\bm{e}_{\rm b}$, and the angular momentum vector, $\bm{l}_{\rm b}$. This binary frame consists of three axes:  $\bm{e}_{\rm b}$,  $\bm{l}_{\rm b}$, and  $\bm{l}_{\rm b} \times  \bm{e}_{\rm b}$. We separate the disc into 300 bins in spherical radius. For discs undergoing coplanar prograde alignment, the radial grid spans from $2a$ to $10a$ and for discs undergoing coplanar retrograde or polar alignment, the radial grid spans from $1.5a$ to $10a$.  Within each bin, we calculate the azimuthally averaged surface density, tilt, longitude of ascending node, and eccentricity. The tilt of an annulus in this disc at radius $R$ relative to the instantaneous binary angular momentum is given by
\begin{equation}
    i(R) = \cos^{-1}(\hat{\bm{l}}_{\rm b}\cdot \hat{\bm{l}}_{\rm d}(r) ),
\end{equation}
where $\hat{\bm{l}}_{\rm b}$ is the unit vector in the direction of the binary angular momentum vector and $\hat{\bm{l}}_{\rm d}(r)$ is the unit vector in the direction of the disc momentum vector. The inclination of the disc is then calculated as the density weighted average of the inclination given by
\begin{equation}
    i = \frac{\int_{r_{\rm in}}^{r_{\rm out}} 2\pi r \Sigma(r)i(r) \,dr}{M_{\rm tot}},
\end{equation}
where $\Sigma (r)$ represents the surface density at radius $r$, $i(r)$ denotes the inclination at radius $r$, and $M_{\rm tot}$ is the total mass of the disc at the specified time. Note, if the disc breaks, an average inclination without being density weighted would become meaningless. The longitude of the ascending node for the disc is given by
\begin{equation}
    \phi(r) = \tan^{-1} \bigg( \frac{\hat{\bm{l}}_{\rm d}(r) \cdot (\hat{\bm{l}}_{\rm b}\times \hat{\bm{e}}_{\rm b})}{\hat{\bm{l}}_{\rm d}(r)\cdot \hat{\bm{e}}_{\rm b}}    \bigg).
\end{equation}
Similarly, $\phi$ is calculated as a density weight average of the longitude of the ascending node given by
\begin{equation}
    \phi = \frac{\int_{r_{\rm in}}^{r_{\rm out}} 2\pi r \Sigma(r)\phi(r) \,dr}{M_{\rm tot}}.
\end{equation}
Lastly, the disc eccentricity at radius $r$ is calculated by
\begin{equation}
    e(r) = \sqrt{\hat{\bm{e}}_{\rm d}(r)\cdot \hat{\bm{e}}_{\rm d}(r)},
\end{equation}
where $\hat{\bm{e}}_{\rm d}(r)$ is the unit vector in the direction of the disc eccentricity vector. The density weighted average of the disc eccentricity is then
\begin{equation}
    e = \frac{\int_{r_{\rm in}}^{r_{\rm out}} 2\pi r \Sigma(r)e(r) \,dr}{M_{\rm tot}},
\end{equation}

% Since we simulate a low disc mass, the tilt, $i$, is defined as the angle between the initial angular momentum vector of the binary (the $z$-axis) and the angular momentum vector of the disc. The longitude of ascending node, $\phi$, is measured relative to the $x$-axis (the initial binary eccentricity vector).

The accretion rate onto each binary component is computed directly from the simulations as the binary is live and it is allowed to accrete gas with time.

\section{Results}
\label{sec::results}

\subsection{Circumbinary disc alignment}

A misaligned disc around an eccentric binary can undergo circulating or librating orbits. The minimum tilt for libration is given by
\begin{equation}
    i_{\rm min} = \arccos\bigg[ \frac{\sqrt{5}e_{\rm b0}\sqrt{4e_{\rm b0}^2 - 4j_0^2(1-e_{\rm b0}^2)+1}-2j_0(1-e_{\rm b0}^2)}{1+4e_{\rm b0}^2} \bigg]
\end{equation}
\citep{Martin2019}, where $j_0 = J_{\rm d0}/J_{\rm b0}$ is the initial angular momentum ratio of the disc to the binary, and $e_{\rm b0}$ is the initial binary eccentricity. The initial angular momentum of the binary is given by
\begin{equation}
    J_{\rm b0} = \mu\sqrt{G(M_1 + M_2)a_{\rm b0}(1-e_{\rm b0}^2)},
\end{equation}
where $\mu $ is the reduced mass and the initial angular momentum of the disc is given by
\begin{equation}
    J_{\rm d0} = \int_{r_{\rm in}}^{r_{\rm out}} 2\pi r^3\Sigma_0(r)\Omega dr,
\end{equation}
where $\Omega = \sqrt{GM/r^3}$ is the Keplerian angular velocity. For our hydro simulations, we have $j_0 \sim 0.0106$ and $e_{\rm b0} = 0.5$, which gives $i_{\rm min} \sim 38^\circ$ or $\sim 142^\circ$ for nodal librating orbits. Therefore, for $i_0 \lesssim 38^\circ$ the disc will evolve to a prograde coplanarity configuration, while for $i_0 \gtrsim 142^\circ$, the disc will evolve to a retrograde coplanarity configuration. For $38^\circ \lesssim i_0 \lesssim 142^\circ$, the disc will undergo nodal libration. Note that for higher binary eccentricity, the minimum inclination angle necessary%needed 
for librating solutions becomes smaller.
%However,  \cite{Martin2018} found that the minimum tilt is slightly higher for a disc than a test particle.

The left panel in Fig.~\ref{fig::tilt} illustrates the average disc tilt as a function of time in binary orbital periods.  The different shades of green curves represent discs undergoing prograde coplanar alignment, while the different shades of blue curves denote discs undergoing retrograde coplanar alignment. The different shades of red curves depict discs undergoing polar alignment (librating orbits). In each model, tilt oscillations occur due to the torque from the eccentric binary \citep{Smallwood2019}. These oscillations dampen over time as the discs evolve closer to their respective alignment configurations. For our selected disc parameters, the polar alignment timescale is faster than coplanar alignment. The right panel in Fig.~\ref{fig::tilt} illustrates the disc evolution in the $i\cos \phi$--$i\sin \phi$ phase space. As mentioned earlier, depending on the initial disc inclination, the disc can reside on a circulating or librating orbit. The centers of the upper libration regions correspond to $i=90^\circ$ and $\phi = 90^\circ$, while the center for the lower librating region corresponds to $i=90^\circ$ and $\phi = -90^\circ$. In this study, we focus on discs evolving towards the upper libration region. The distance from the center of the circulating or librating regions to any point on the phase diagram curve indicates the tilt of the disc. The more "oval-shaped" the phase diagram, the larger the amplitude of the disc tilt oscillations.

\begin{figure} 
\centering
\includegraphics[width=1\columnwidth]{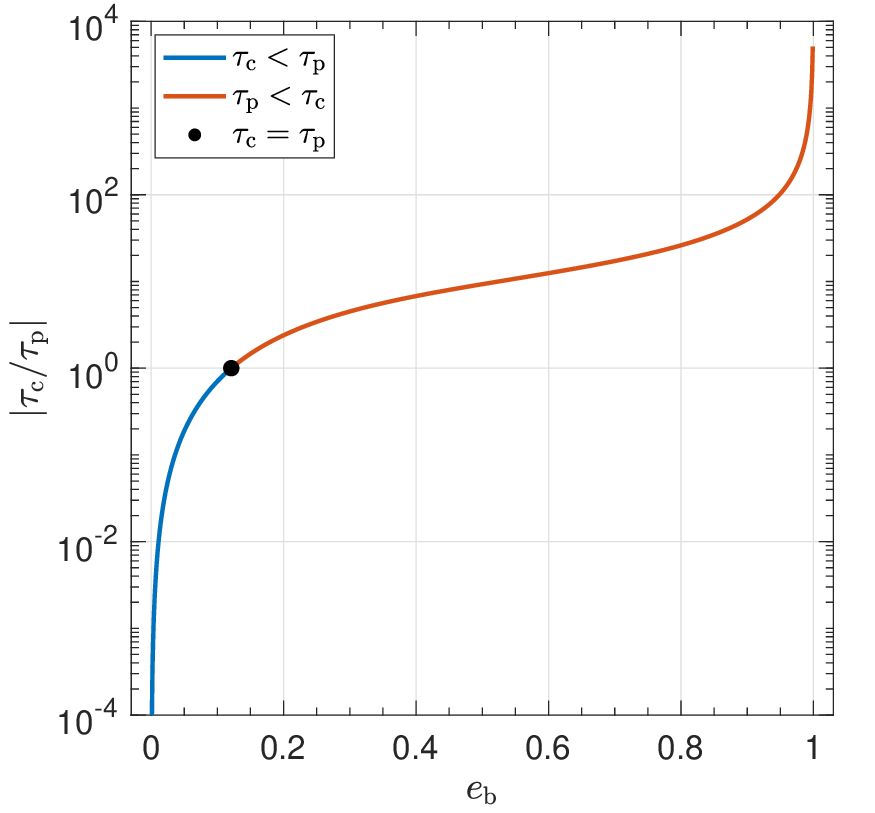}
\caption{The ratio of the coplanar alignment timescale to the polar alignment timescale, $\tau_{\rm c}/\tau_{\rm p}$ (from Eq.~(\ref{eq::ratio})), as a function of binary eccentricity.  The blue curve represents $\tau_{\rm c}/\tau_{\rm p} < 1$ and the red curve represents $\tau_{\rm p}/\tau_{\rm c} < 1$. The black dot denotes when $\tau_{\rm c}/\tau_{\rm p} = 1$.}
\label{fig::alignment}
\end{figure}

\begin{figure*} 
\begin{center}
    \includegraphics[width=1\columnwidth]{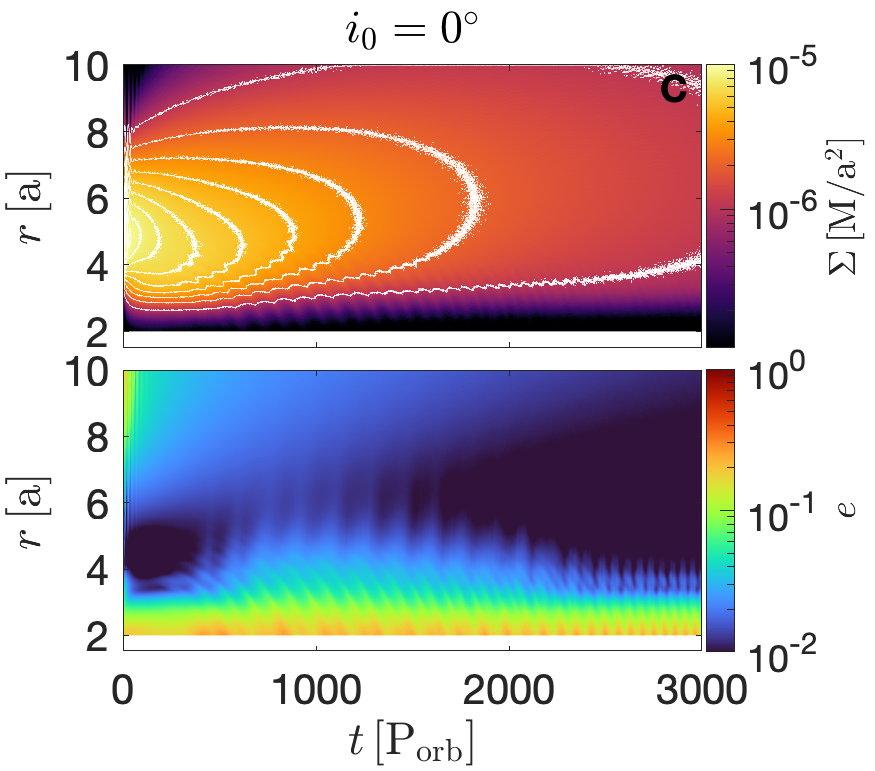}
\end{center}
\includegraphics[width=0.5\columnwidth]{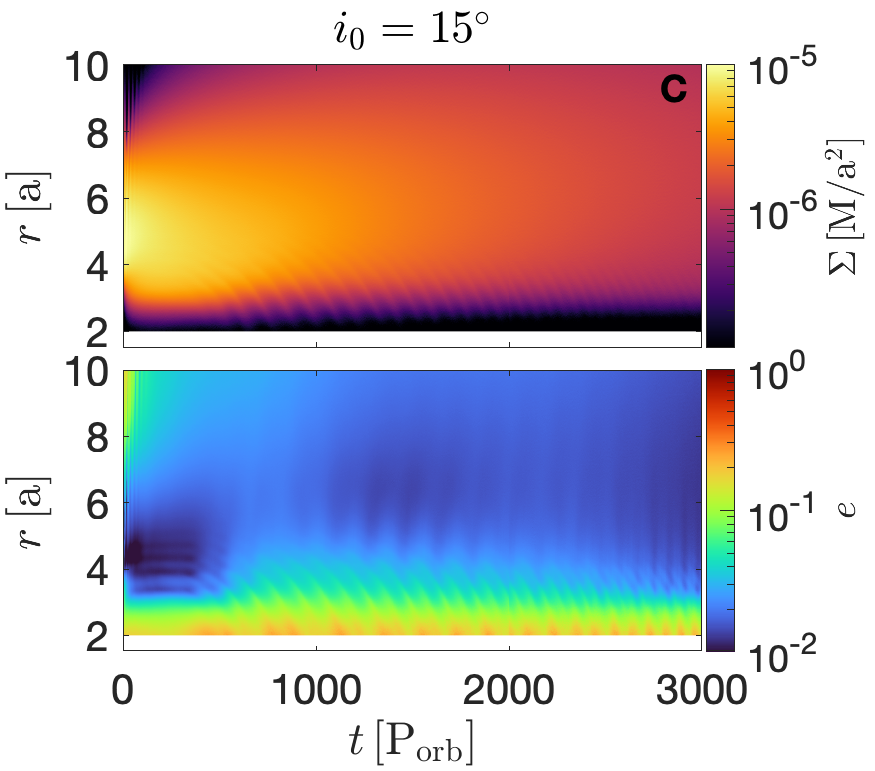}
\includegraphics[width=0.5\columnwidth]{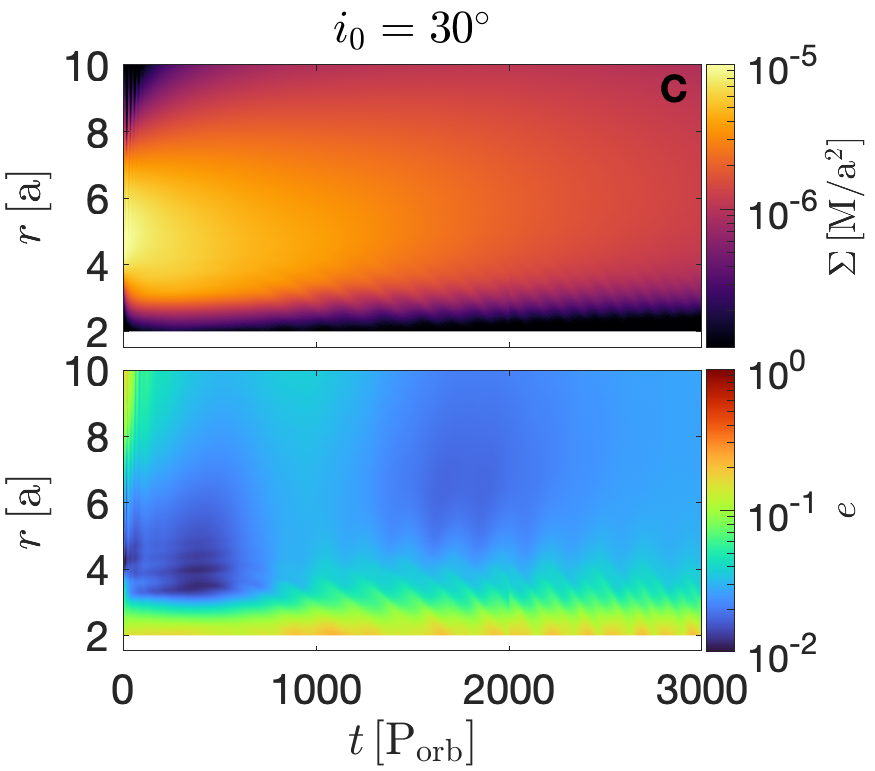}
\includegraphics[width=0.5\columnwidth]{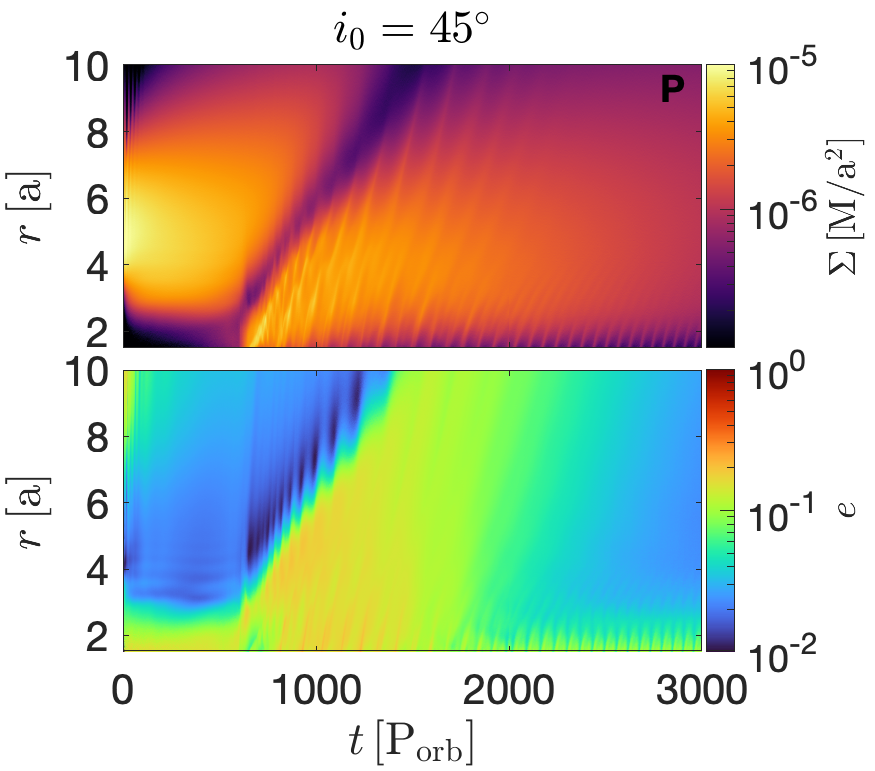}
\includegraphics[width=0.5\columnwidth]{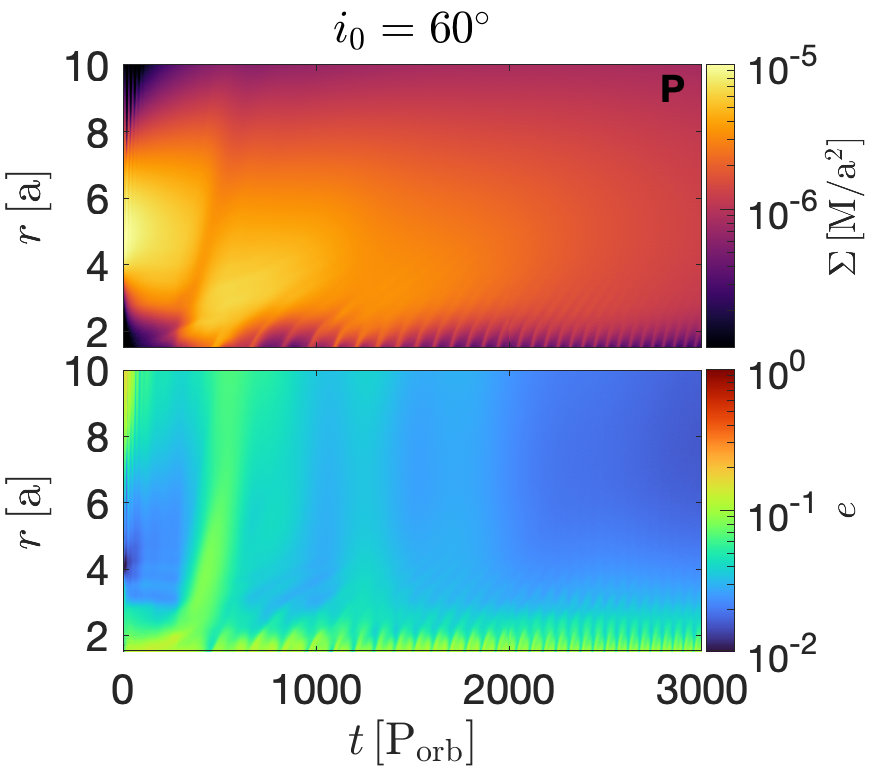}
\includegraphics[width=0.5\columnwidth]{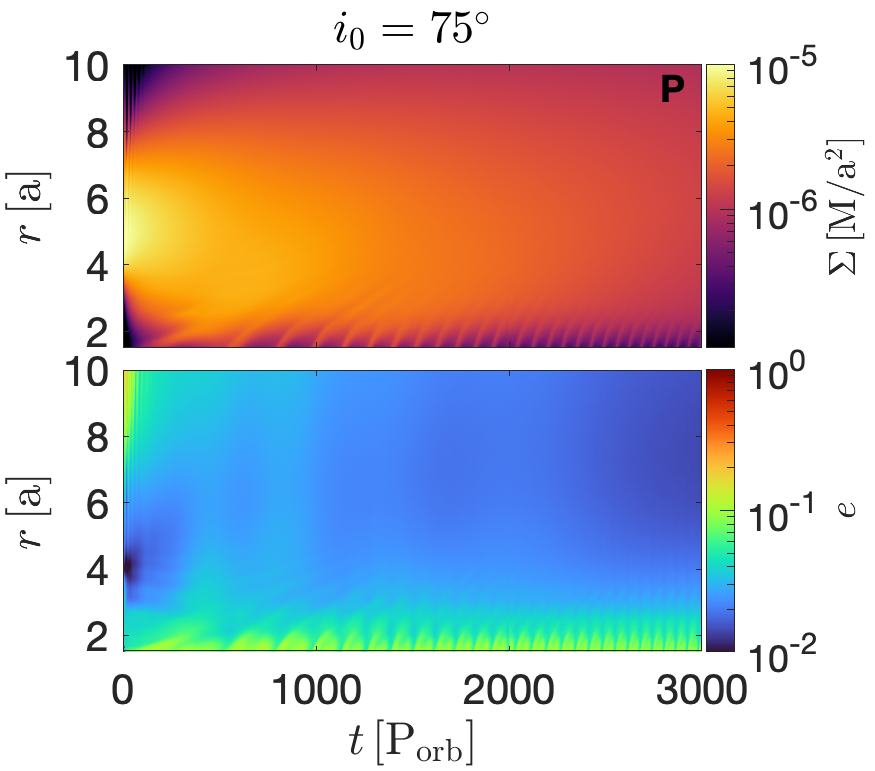}
\includegraphics[width=0.5\columnwidth]{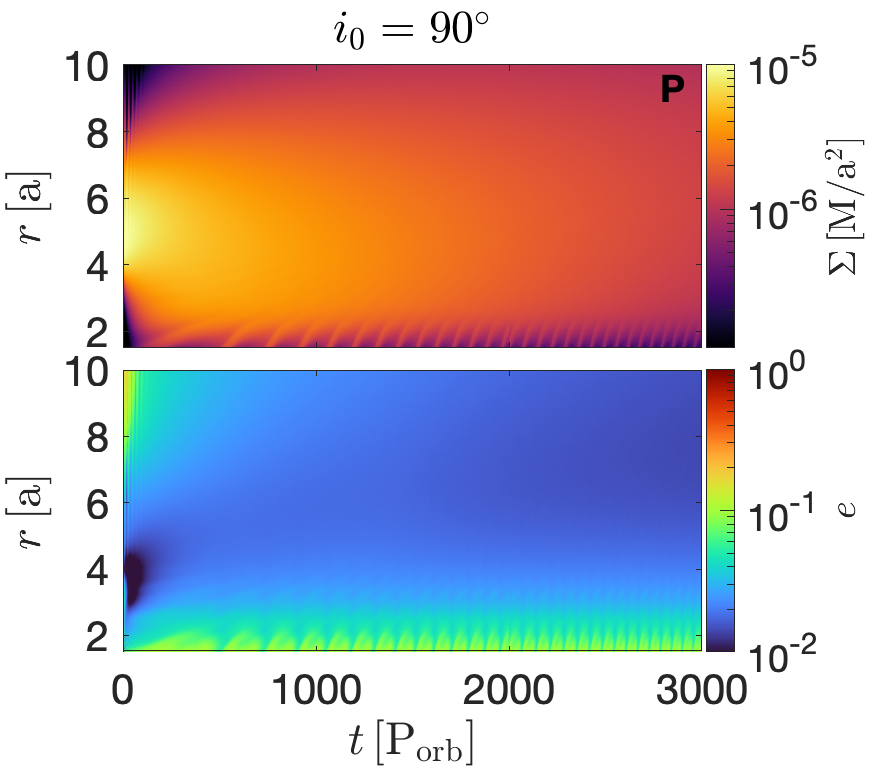}
\includegraphics[width=0.5\columnwidth]{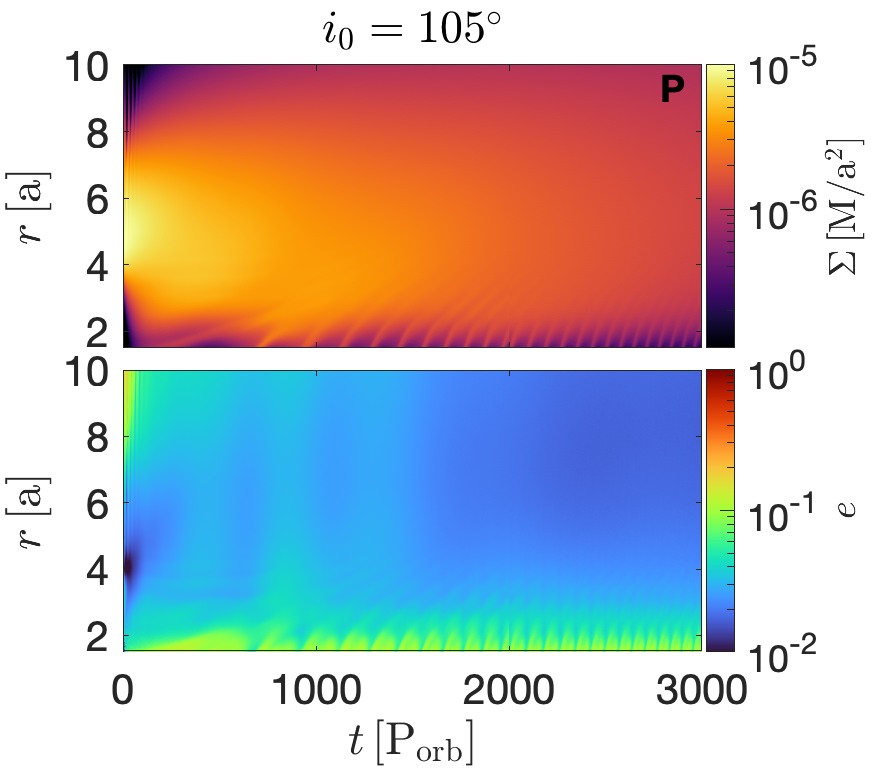}
\includegraphics[width=0.5\columnwidth]{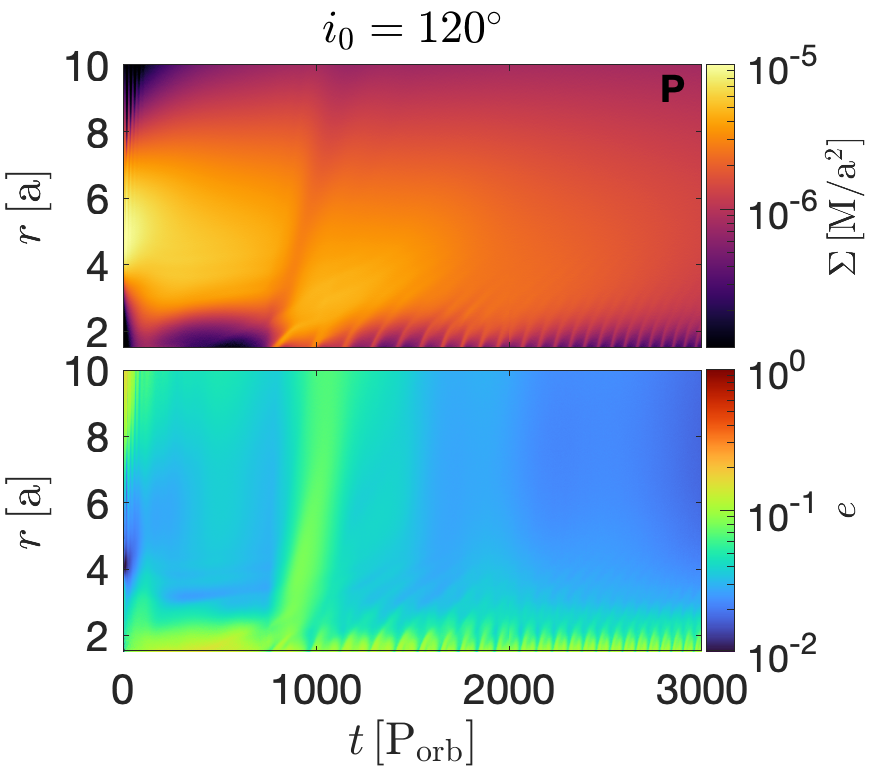}
\includegraphics[width=0.5\columnwidth]{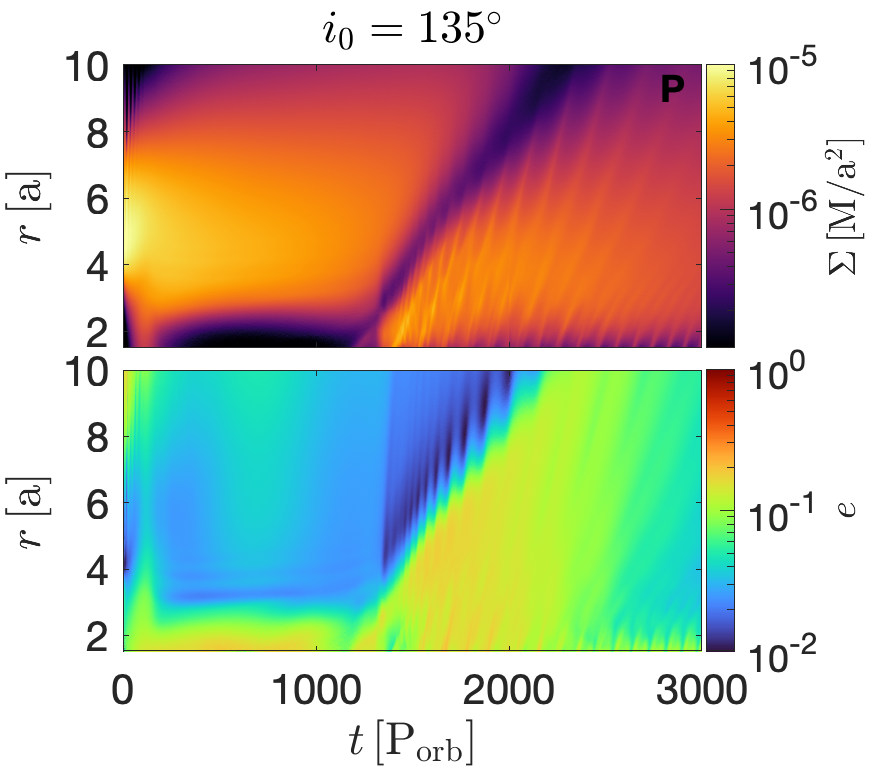}
\includegraphics[width=0.5\columnwidth]{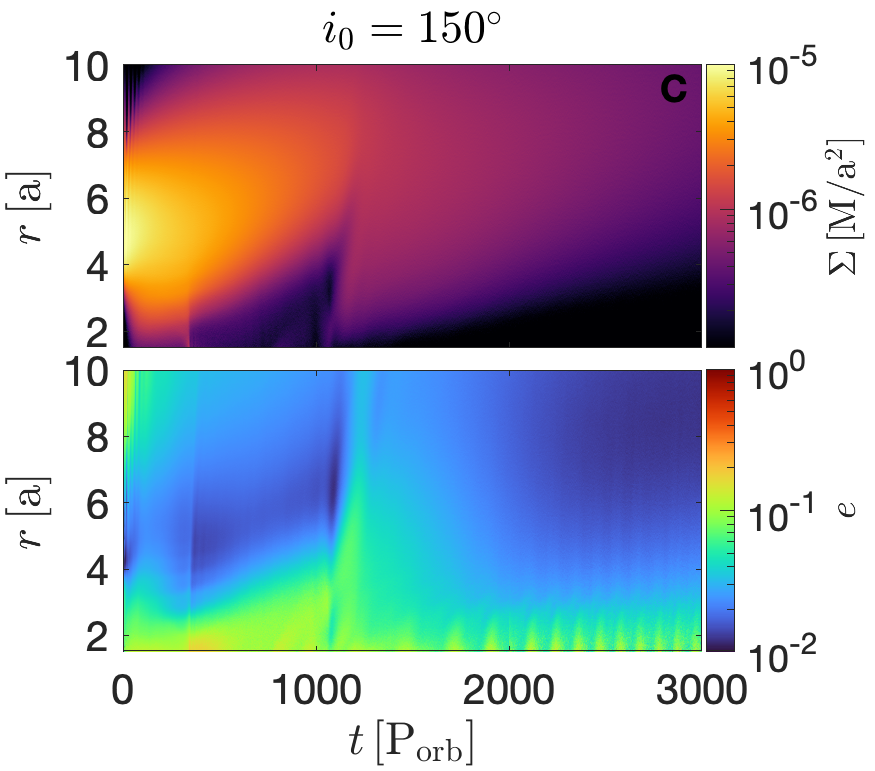}
\includegraphics[width=0.5\columnwidth]{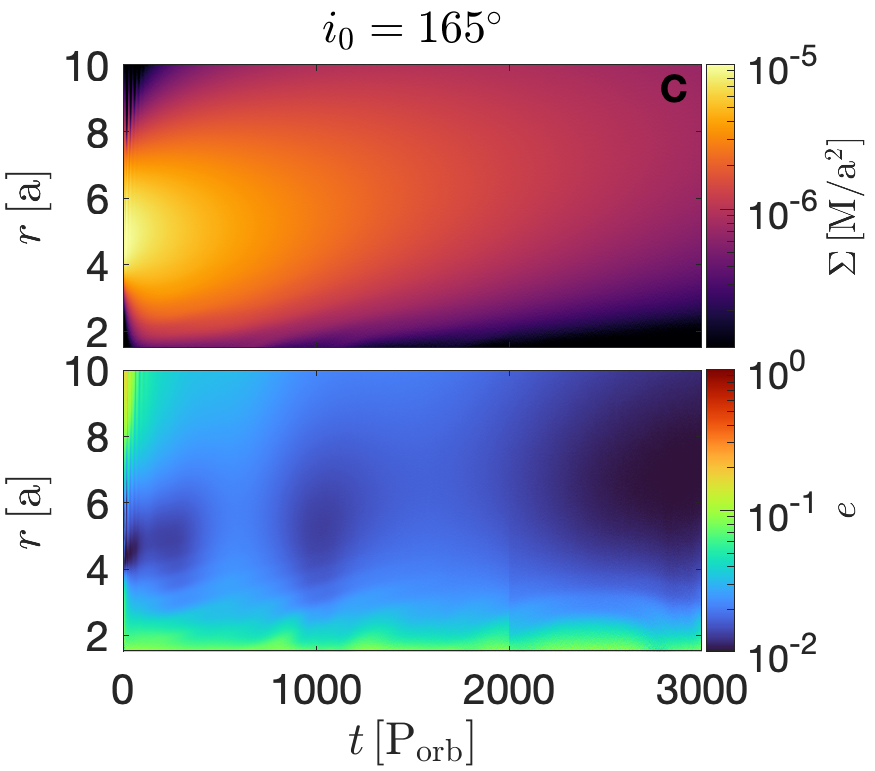}
\includegraphics[width=0.5\columnwidth]{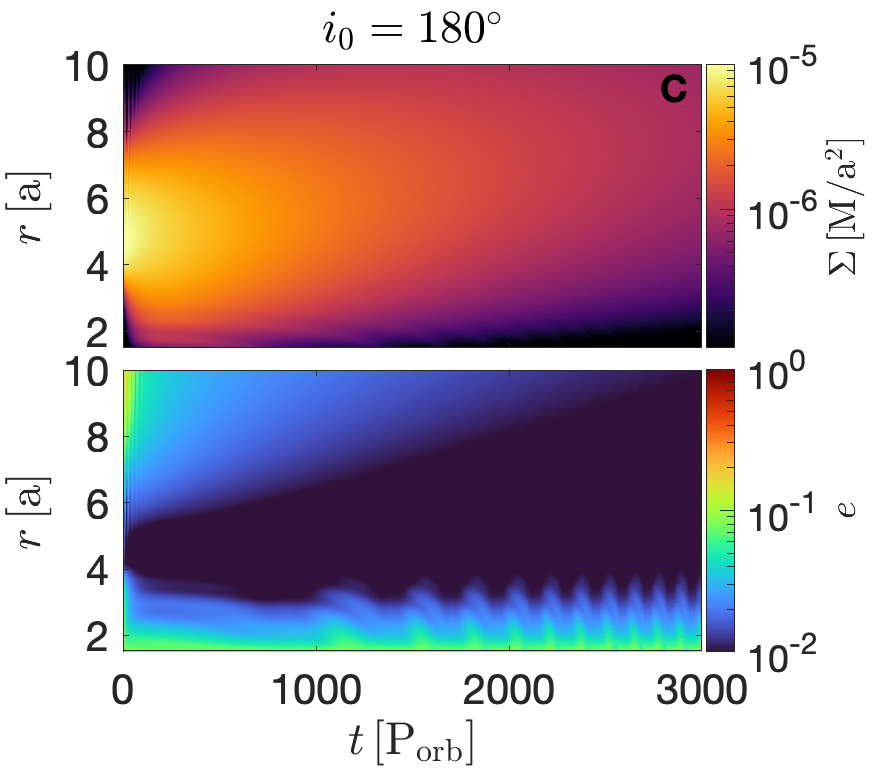}
\caption{The disc surface density $\Sigma$ (upper subpanel), and eccentricity $e$ (lower subpanel), as a function of time in units of initial binary orbital period, $P_{\rm orb}$, for different initial misalignment of the circumbinary disc (given by the titles).  We show 15 contour levels of the surface density in the enlarged panel. The letters "C" or "P" denote whether the disc is undergoing coplanar or polar alignment, respectively. Note that in the $45^{\circ}$ and $135^{\circ}$ the upper panel clearly shows a break, i.e. a discontinuity in the surface density profile. }
\label{fig::ecc}
\end{figure*}

The process driving alignment between the accretion disc and the binary angular momenta stems from the inherent viscosity within the disc. This viscosity effectively reduces inclination oscillations over a timescale proportional to $\alpha^{-1}$, thereby dissipating the warp \citep{Papaloizou1995,Lubow2000,Lubow2018}. 
The linear theory of warped discs describes the evolution of the warp in the regime of small inclination angles.% the evolution of a disc that remains nearly flat. 
Within the linear warp propagation theory, the warp is dissipated and the circumbinary disc aligns on a timescale
\begin{equation}
    \tau = \frac{1}{\alpha}\bigg( \frac{H}{r}\bigg)^2 \frac{\Omega_{\rm b}}{\Omega_{\rm d}^2},
    \label{eq::tau}
\end{equation}
where $\alpha$ is the disc viscosity, $H/r$ is the disc aspect ratio, and $\Omega_{\rm b} = \sqrt{G(M_1 + M_2)/a^3}$ is the binary angular frequency. $\Omega_{\rm d}$ is the global disc precession frequency, which differs depending on coplanar or polar alignment, and is given by
\begin{equation}
  \Omega_{\rm d}=\begin{cases}
    -\frac{3}{4}\sqrt{1+3e_{\rm b}^2-4e_{\rm b}^4}\frac{M_1M_2}{M^2}\left<\left( \frac{a}{r}\right)^{7/2}\right>\Omega_{\rm b}, & \text{coplanar},\\
    \frac{3\sqrt{5}}{4}e_{\rm b}\sqrt{1+4e_{\rm b}^2}\frac{M_1M_2}{M^2}\left<\left( \frac{a}{r}\right)^{7/2}\right>\Omega_{\rm b}, & \text{polar}
  \end{cases}
  \label{eq::omega}
\end{equation}
\citep{Lubow2018,Smallwood2019},
where
\begin{equation}
    \left<\left( \frac{a}{r}\right)^{7/2}\right>=\frac{\int_{r_{\rm c/p}}^{r_{\rm out}}\Sigma r^3 \Omega (a/r)^{7/2}\,dr }{\int_{r_{\rm c/p}}^{r_{\rm out}} \Sigma r^3 \Omega \, dr},
    \label{eq::ar}
\end{equation}
and where $\Omega=\sqrt{G(M_1+M_2)/r^3}$ is the Keplerian angular frequency. The lower bound for the integrals, $r_{\rm c}$ and $r_{\rm p}$, corresponds to the inner edge for a coplanar and polar discs, respectively. Note that if the disc breaks, the constituent discs will precess at different rates and Eq.~\ref{eq::ar} cannot be applied. Using Eqs.~(\ref{eq::tau}) and~(\ref{eq::omega}), we calculate the ratio of the coplanar alignment timescale to the polar alignment timescale to be
% \begin{equation}
%     \frac{\tau_{\rm c}}{\tau_{\rm p}} = \frac{5e_{\rm b}^2}{-1 + e_{\rm b}^2}.
%     \label{eq::ratio}
% \end{equation}
\begin{equation}
    \frac{\tau_{\rm c}}{\tau_{\rm p}} = \frac{5e_{\rm b}^2 (1+4e_{\rm b}^2)}{1 + 3e_{\rm b}^2 -4e_{\rm b}^4} \bigg(\frac{10-r_{\rm c}}{10-r_{\rm p}}\bigg)^2 \bigg( \frac{-625r_{\rm p}^{5/2}+\sqrt{10}}{-625r_{\rm c}^{5/2}+\sqrt{10}} \bigg)^2 \bigg(  \frac{r_{\rm c}}{r_{\rm p}}\bigg)^5.
    \label{eq::ratio}
\end{equation}
Assuming that the disc parameters ($\Sigma$, $\alpha$, and $H/r$) are equivalent in the coplanar and polar models, $\tau_{\rm c}/\tau_{\rm p}$ is dependent on $e_{\rm b}$, $r_{\rm c}$, and $r_{\rm p}$. $r_{\rm c}$ can vary with binary eccentricity, and can be estimated from \cite{Artymowicz1994}. $r_{\rm p}$ will not vary significantly with binary eccentricity since the binary tidal torque is weaker for polar-aligning discs. We assume a constant value of $r_{\rm p} = 1.5a$ \citep{Kennedy2019}.
 
Figure~\ref{fig::alignment} shows the ratio of the coplanar alignment timescale to the polar alignment timescale, $\tau_{\rm c}/\tau_{\rm p}$ (from Eq.~(\ref{eq::ratio})), as a function of binary eccentricity. At $e_{\rm b} = 0.12$, the ratio $\tau_{\rm c}/\tau_{\rm p} \approx 1$.  The binary eccentricity value when $\tau_{\rm c}/\tau_{\rm p} \approx 1$ will change if the coplanar and polar discs have different disc parameters ($\Sigma$, $\alpha$, and $H/r$) with respect to one another. Therefore, for $e_{\rm b} < 0.12$, the polar alignment timescale is longer than the coplanar alignment timescale. Conversely, for $e_{\rm b} > 0.12$, the polar alignment timescale is shorter than the coplanar alignment timescale. Given that our simulations have $e_{\rm b} = 0.5$, the discs aligning polar will have a shorter alignment timescale than the simulations modelling coplanar aligning discs (see the left panel in Fig.~\ref{fig::tilt}).  These results are in line with previous work \cite[e.g.,][]{LiD2014,Miranda2015,Martin2017,Martin2019,Cuello2019,Ceppi2023}.

 % Equation (\ref{eq:tbate}) provides a basic approximation, overlooking variations in quantities like $\alpha$ and the disc aspect ratio $H/R$ with radius $R". To make accurate predictions based on linear theory, one typically computes the complex eigenfrequency using the linear tilt evolution equations (e.g., Lubow et al., 2018). However, we refrain from conducting such computations in this context. It's worth noting that the linear model we employ here doesn't consider the evolution of the density distribution, as discussed in Martin and Lubow (2019).

\subsection{Circumbinary disc structure and evolution}

\begin{figure} 
\centering
\includegraphics[width=1\columnwidth]{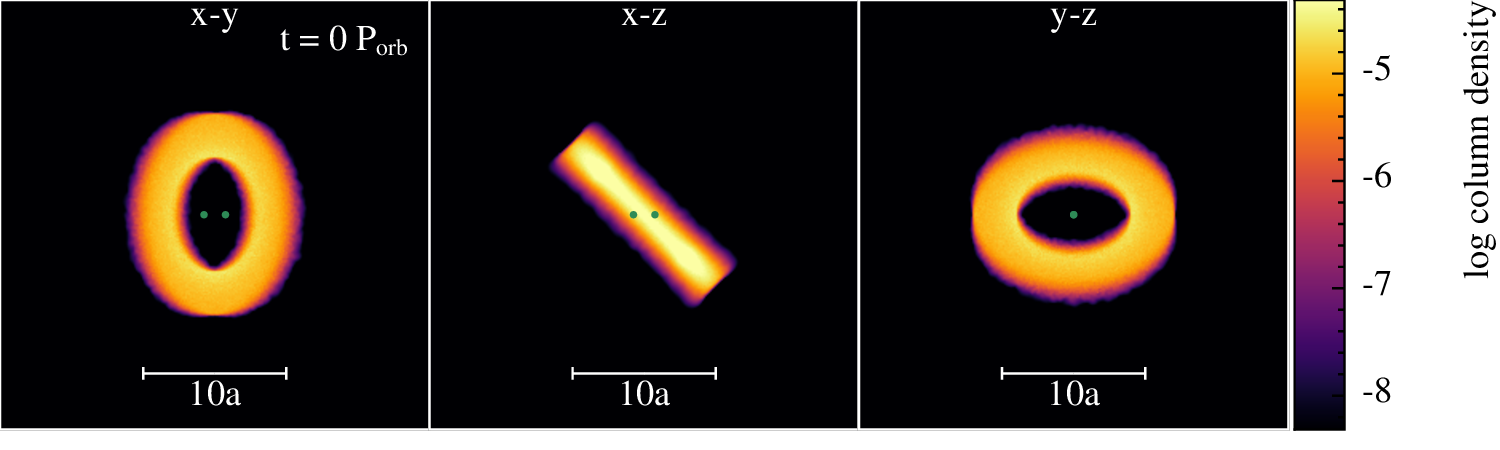}
\includegraphics[width=1\columnwidth]{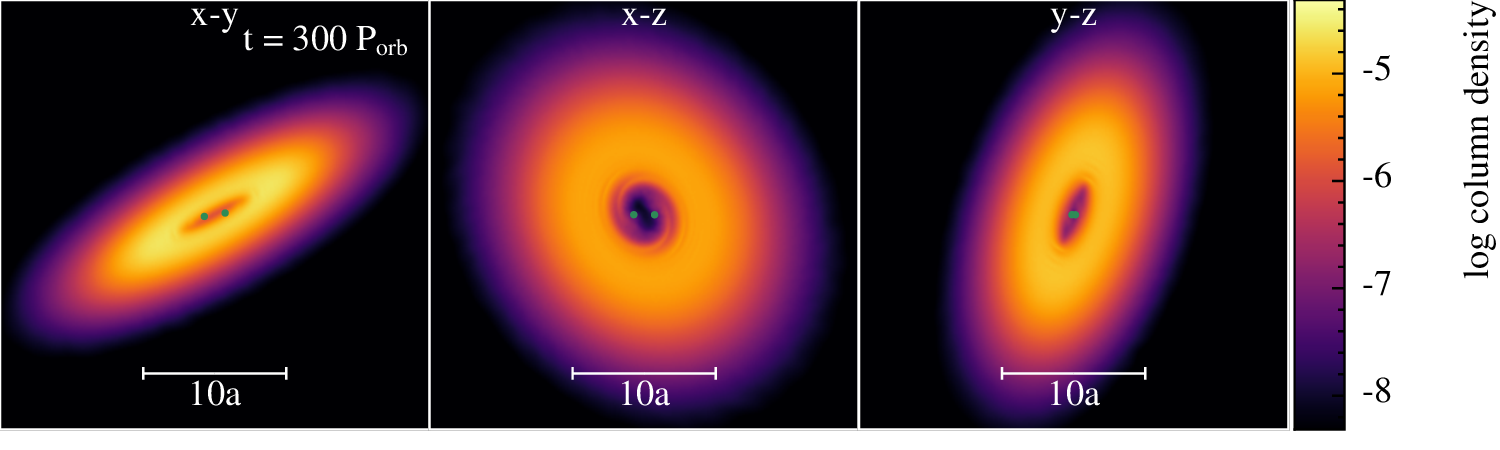}
\includegraphics[width=1\columnwidth]{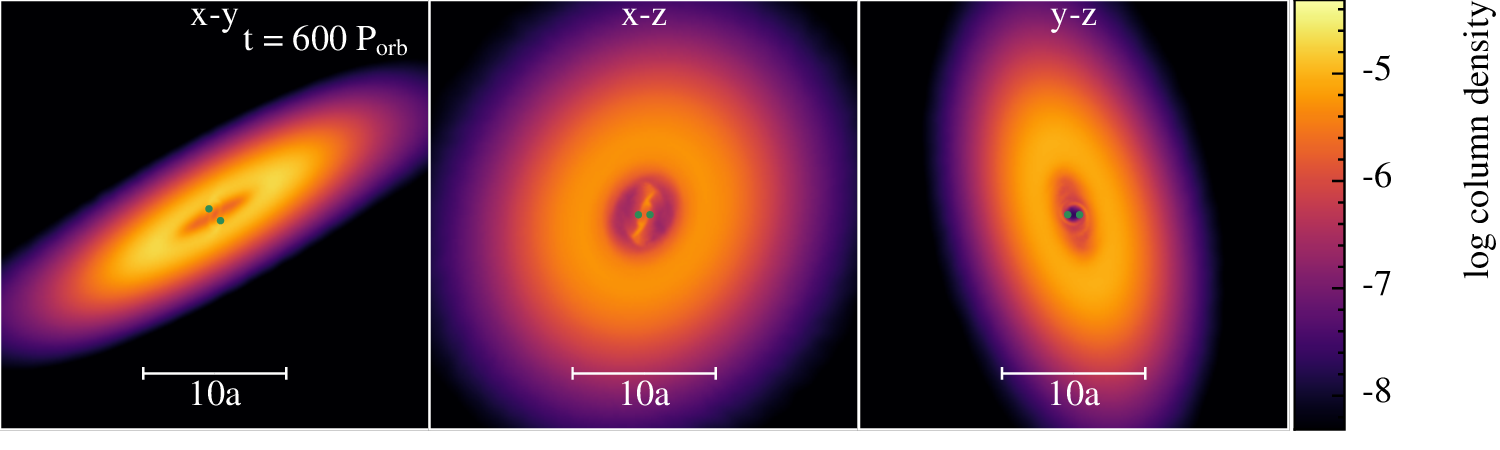}
\includegraphics[width=1\columnwidth]{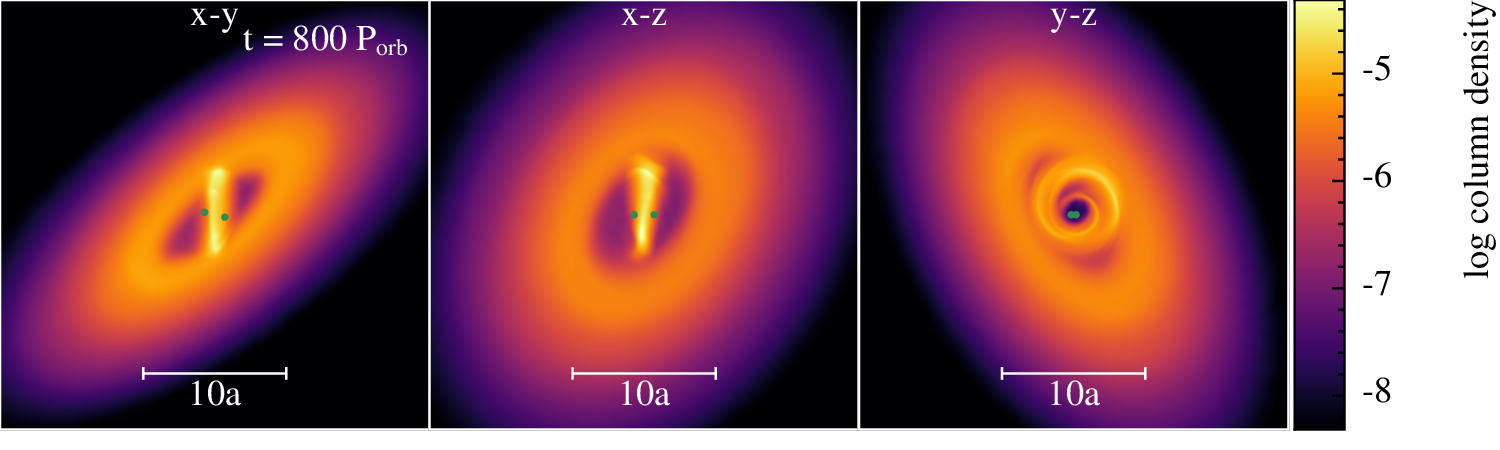}
\includegraphics[width=1\columnwidth]{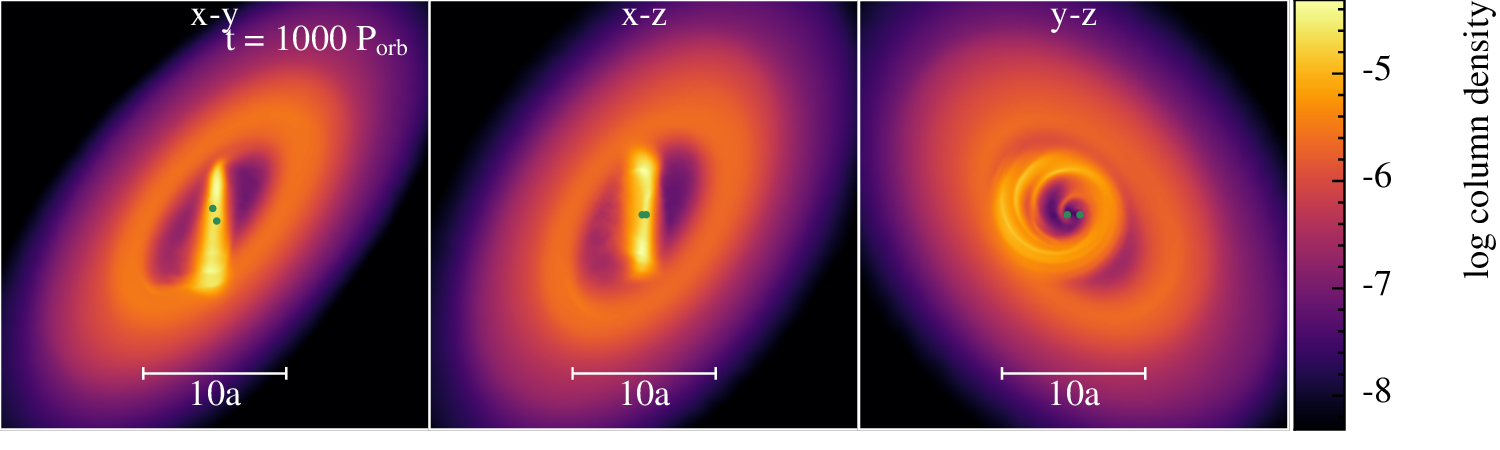}
\includegraphics[width=1\columnwidth]{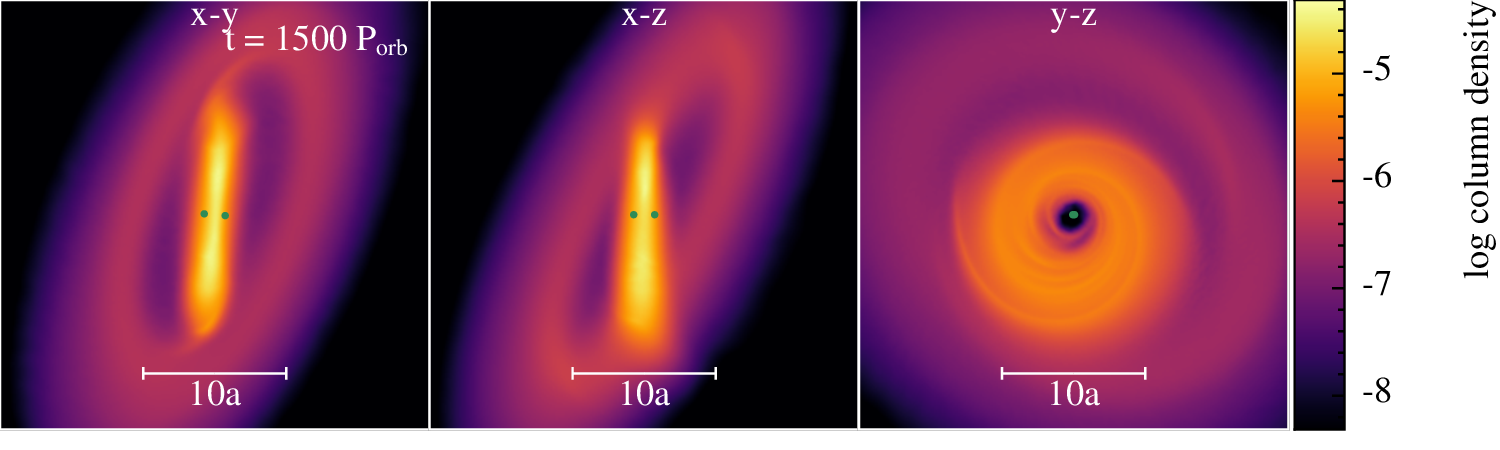}
\includegraphics[width=1\columnwidth]{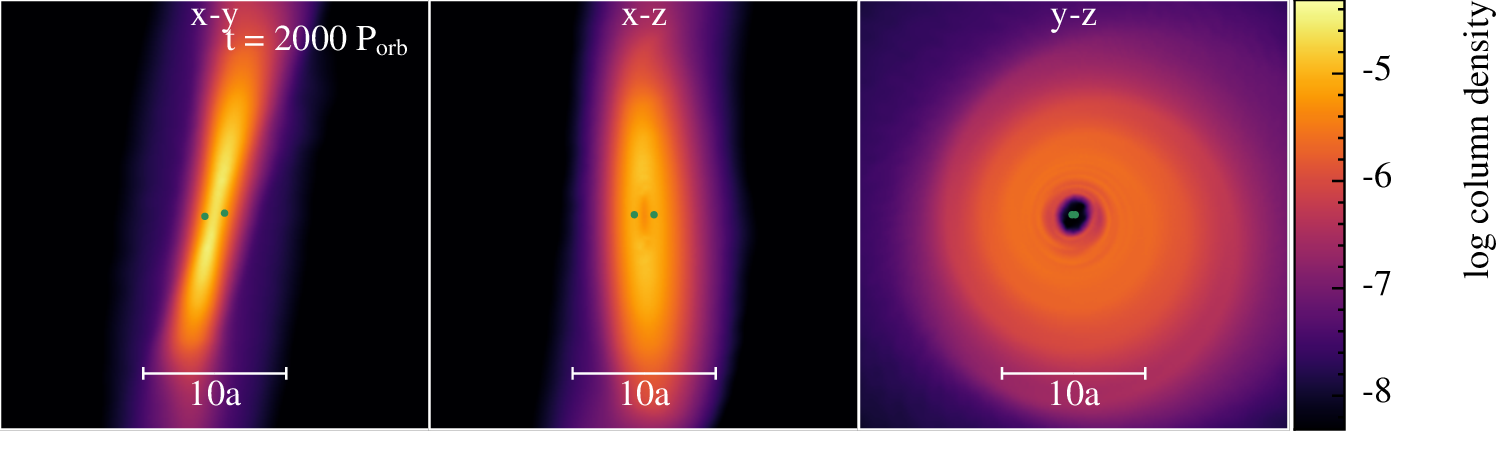}
\includegraphics[width=1\columnwidth]{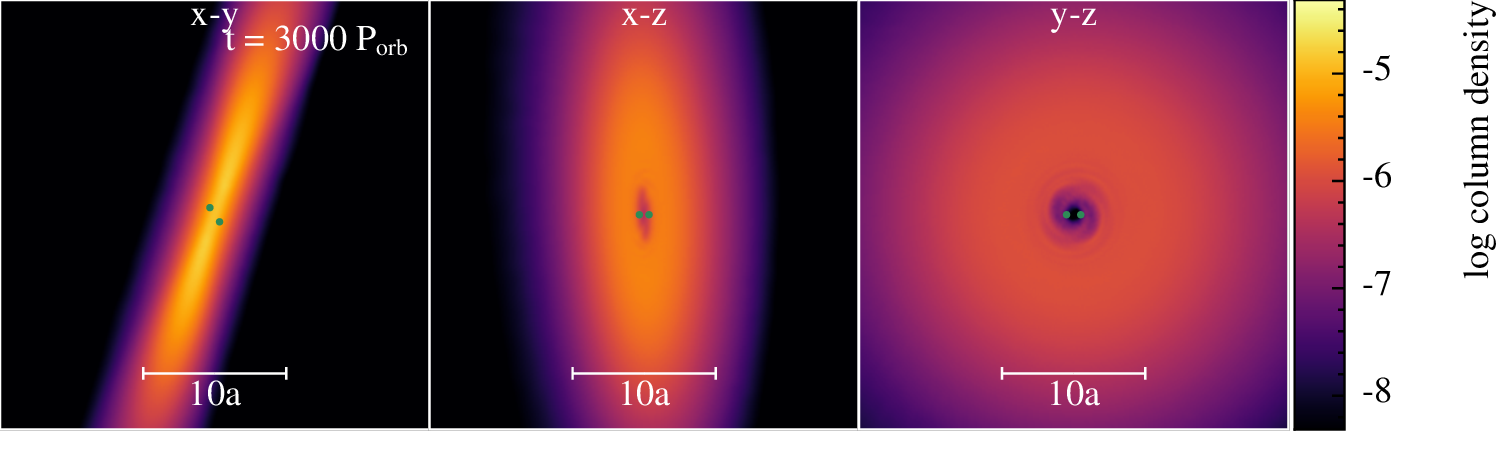}
\caption{The gas surface density for a circumbinary disc initially misaligned by $i_0 = 45^\circ$. The left sub-panel shows the $x$--$y$ plane, viewing down on the binary orbit. The middle sub-panel denotes the $x$--$z$ plane, which shows the initial misalignment between the disc and binary, and the right sub-panel shows the $y$--$z$ plane. We show selected times ranging from $t = 0\, \rm P_{orb}$ to $t = 3000\, \rm P_{orb}$, where $\rm P_{orb}$ is the initial binary orbital period.}
\label{fig::splash_i45}
\end{figure}

\begin{figure} 
\centering
\includegraphics[width=1\columnwidth]{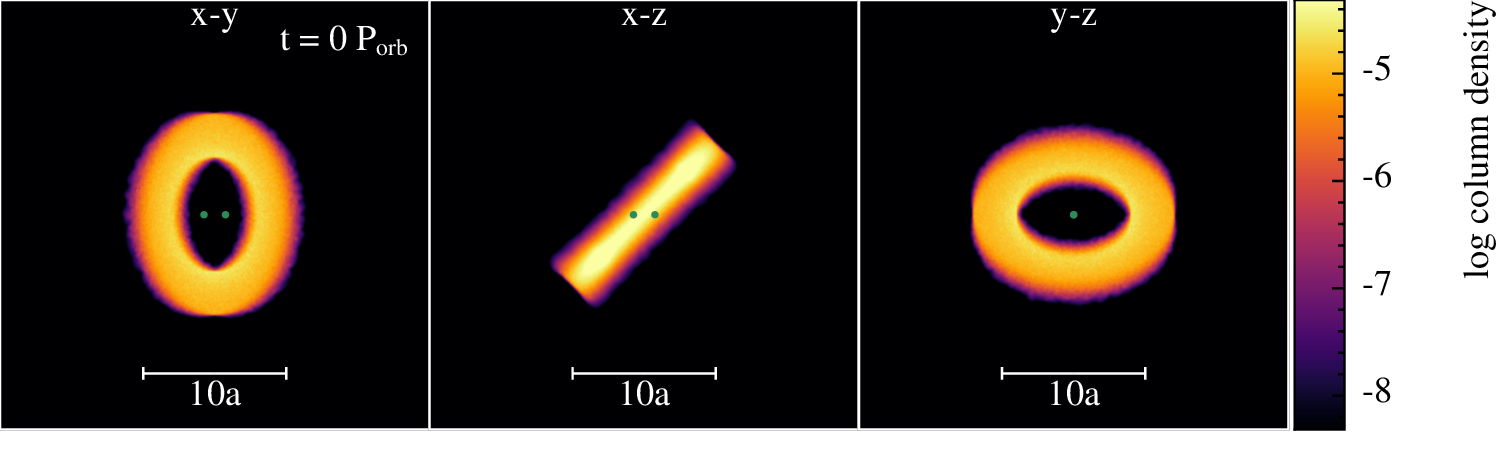}
\includegraphics[width=1\columnwidth]{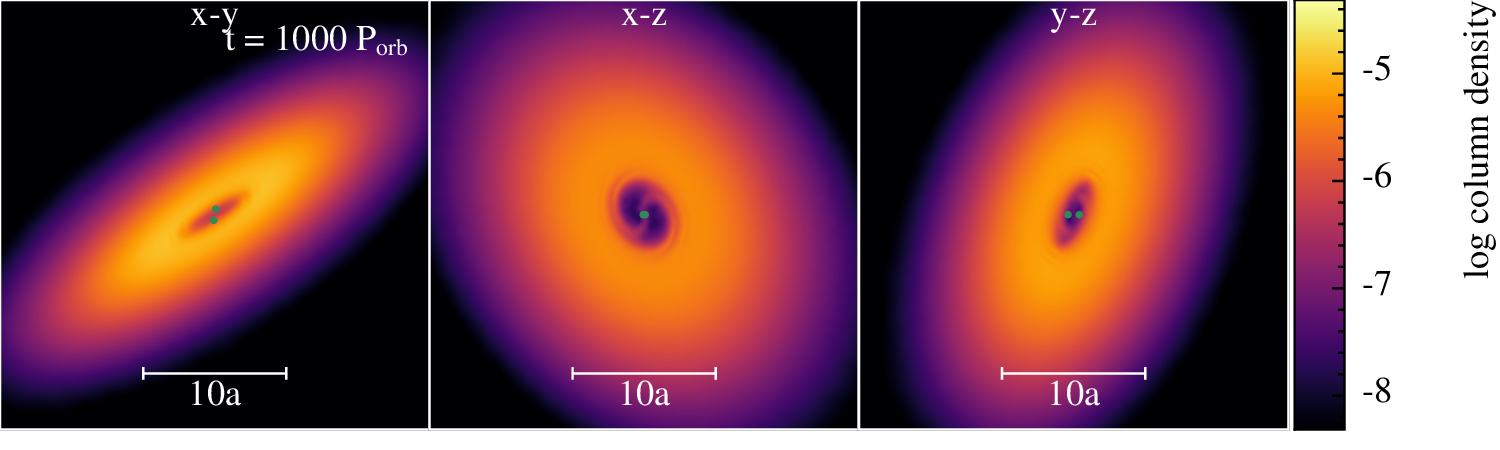}
\includegraphics[width=1\columnwidth]{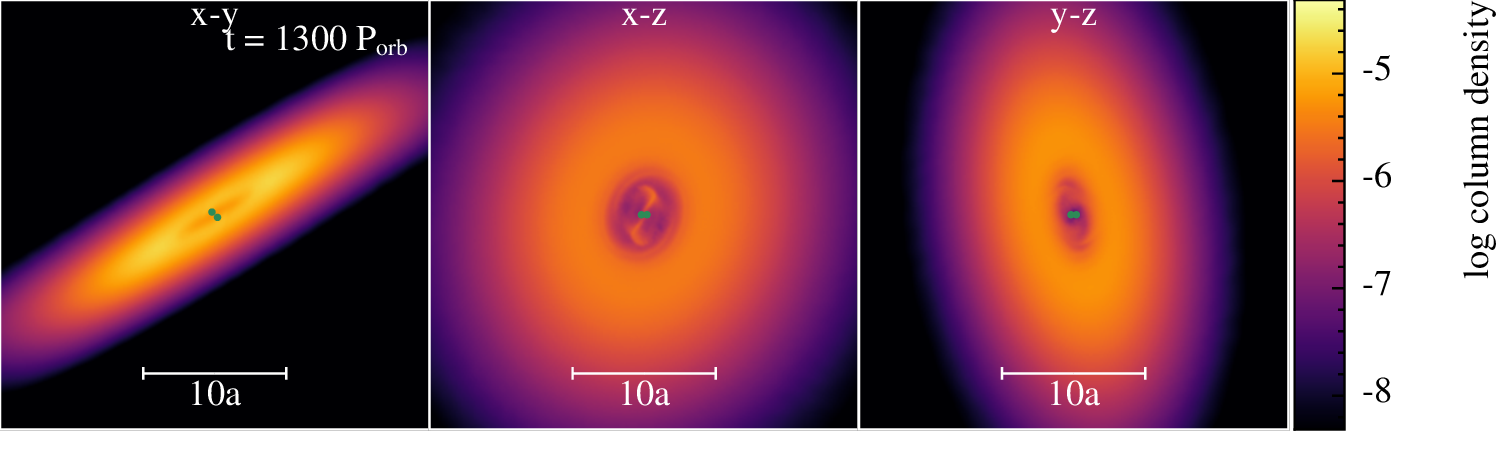}
\includegraphics[width=1\columnwidth]{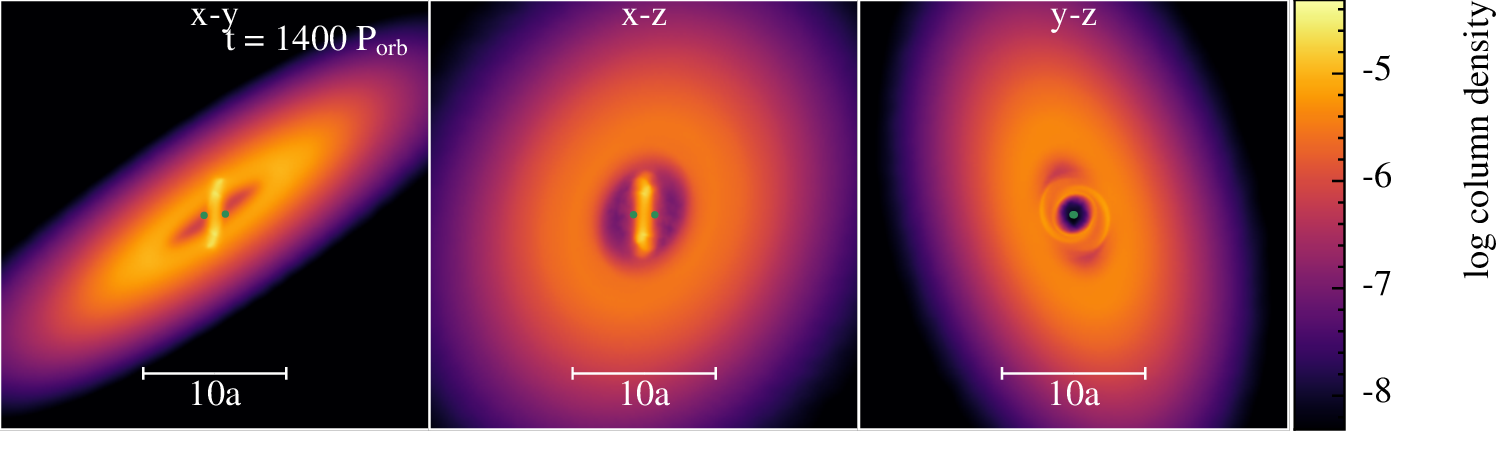}
\includegraphics[width=1\columnwidth]{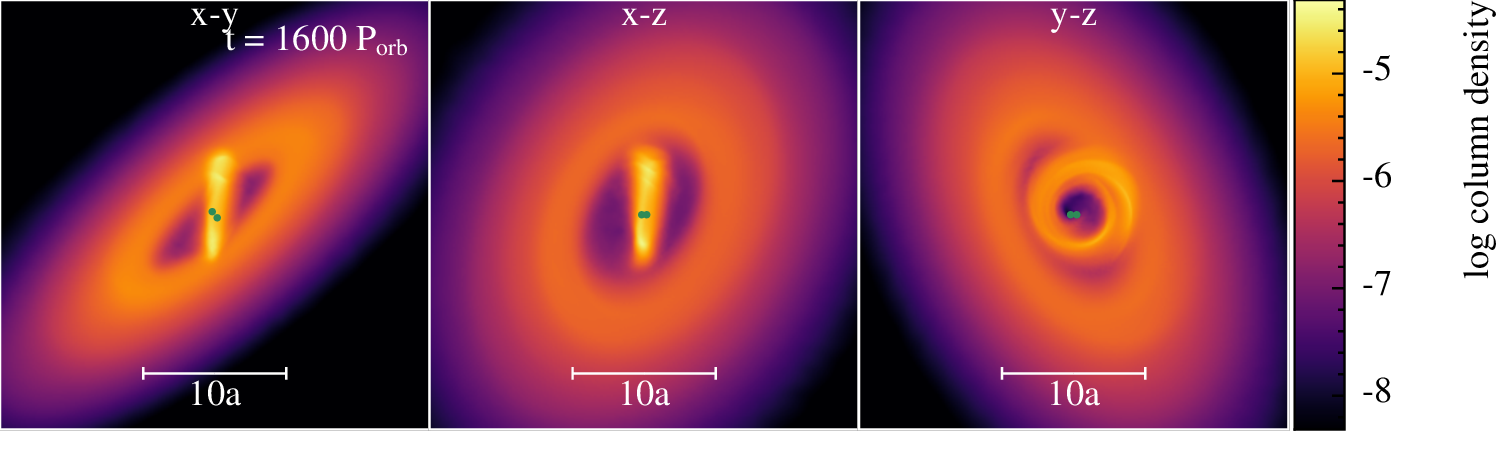}
\includegraphics[width=1\columnwidth]{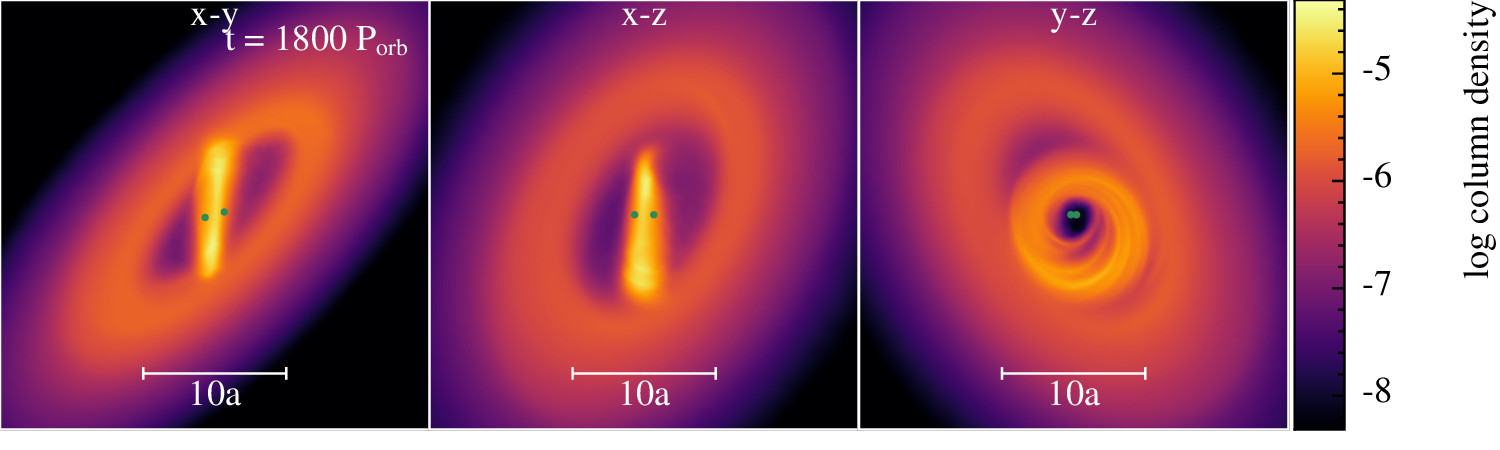}
\includegraphics[width=1\columnwidth]{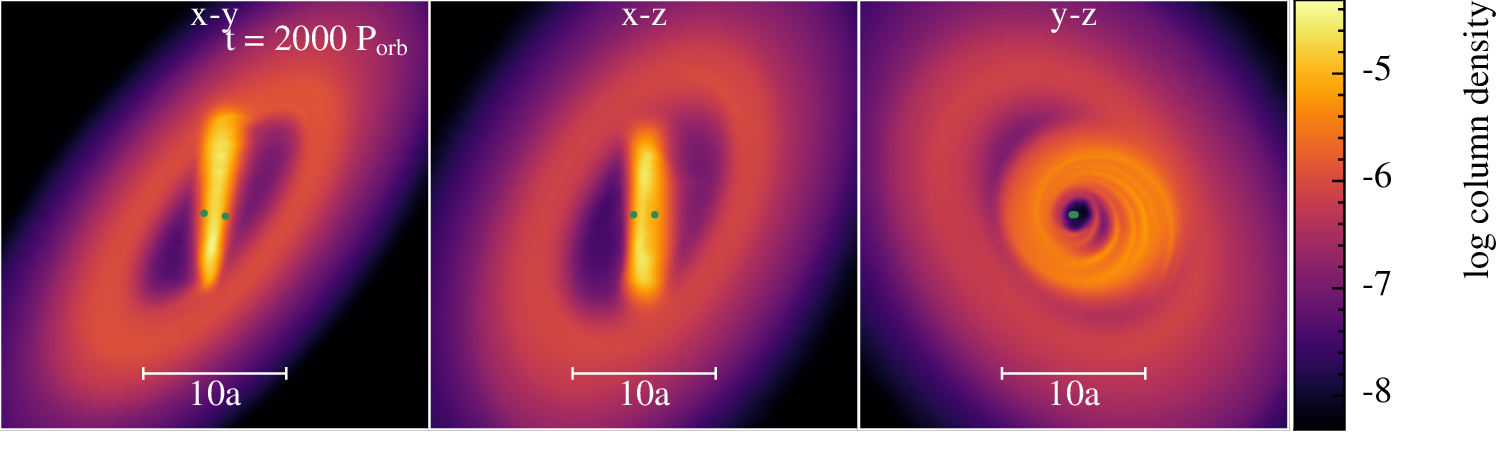}
\includegraphics[width=1\columnwidth]{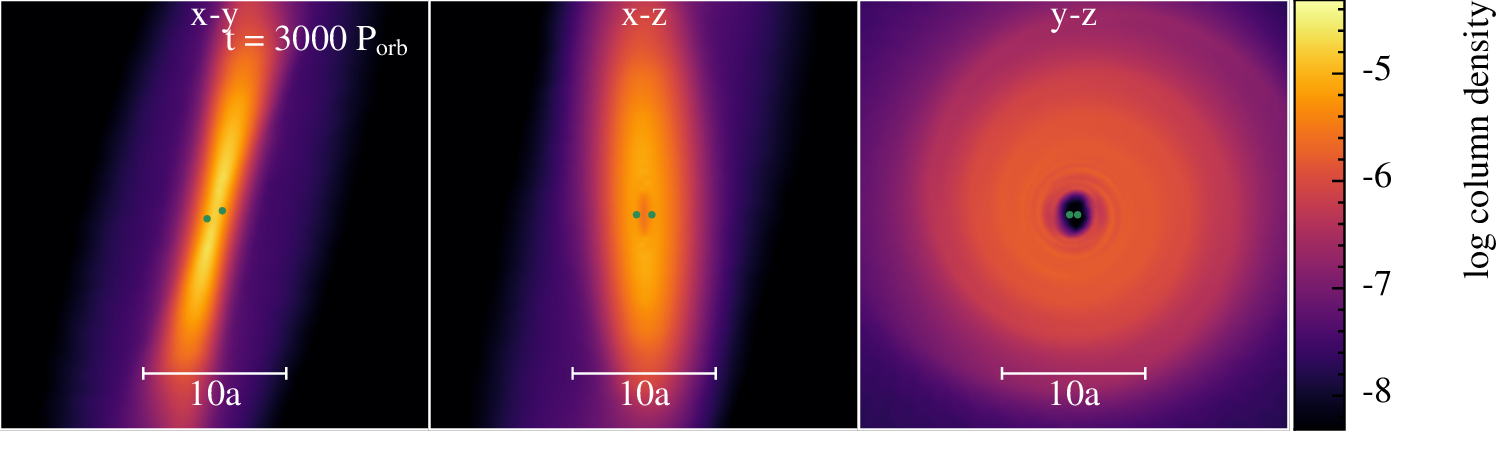}
\caption{Same as Fig.~\ref{fig::splash_i45}, but for a initial circumbinary disc tilt of $i_0 = 135^\circ$.}
\label{fig::splash_i135}
\end{figure}

\begin{figure} 
\centering
\includegraphics[width=\columnwidth]{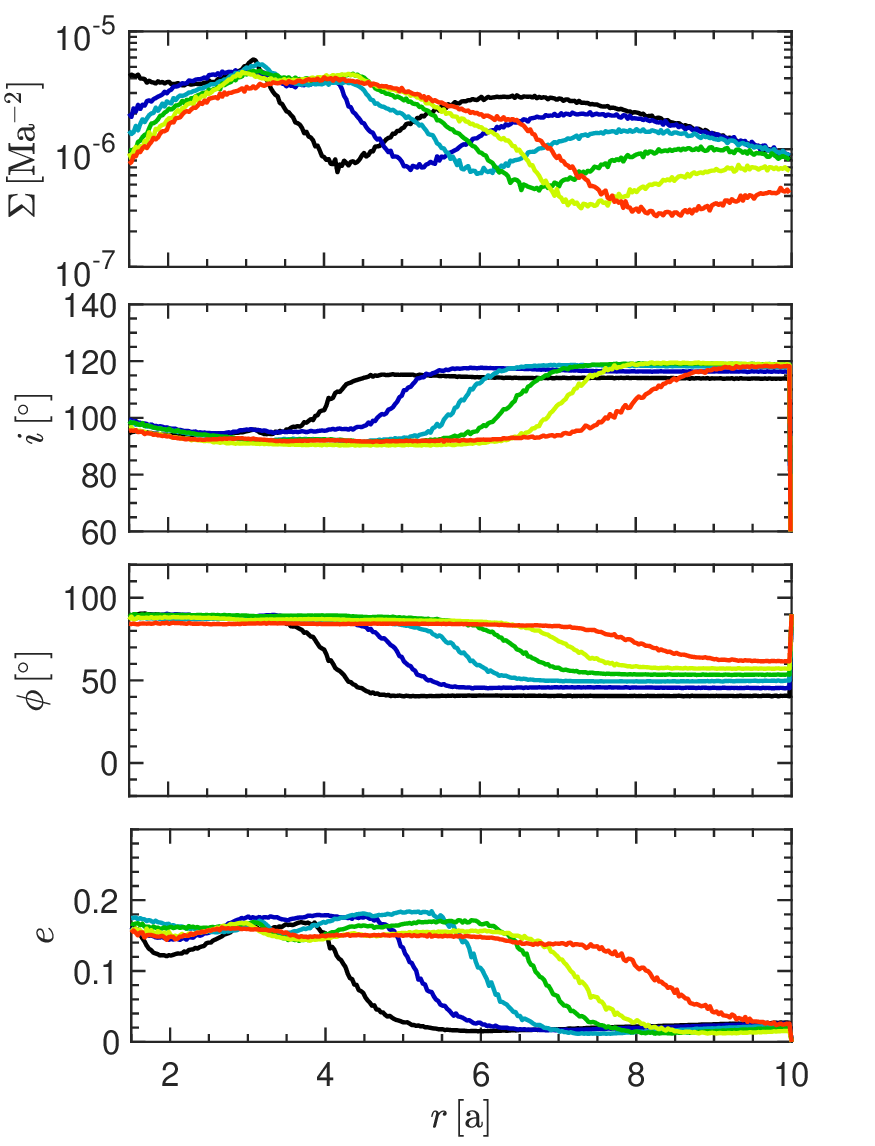}
\caption{The evolution of the surface density, $\Sigma$, tilt, $i$, longitude of the ascending node, $\phi$, and eccentricity, $e$, as a function of radius, $r$, for model of $i_{0}=45^{\circ}$ during the disc-breaking stage. We show selected times beginning with $t = 800\, \rm P_{orb}$ (black), $900\, \rm P_{orb}$ (blue), $1000\, \rm P_{orb}$ (teal), $1100\, \rm P_{orb}$ (green), $1200\, \rm P_{orb}$ (yellow), and $1300\, \rm P_{orb}$ (red). The break in the disc propagates outward as the disc aligns to a polar state.}
\label{fig::i45_disc_params}
\end{figure}

\begin{figure} 
\begin{center}
\includegraphics[width=1\columnwidth]{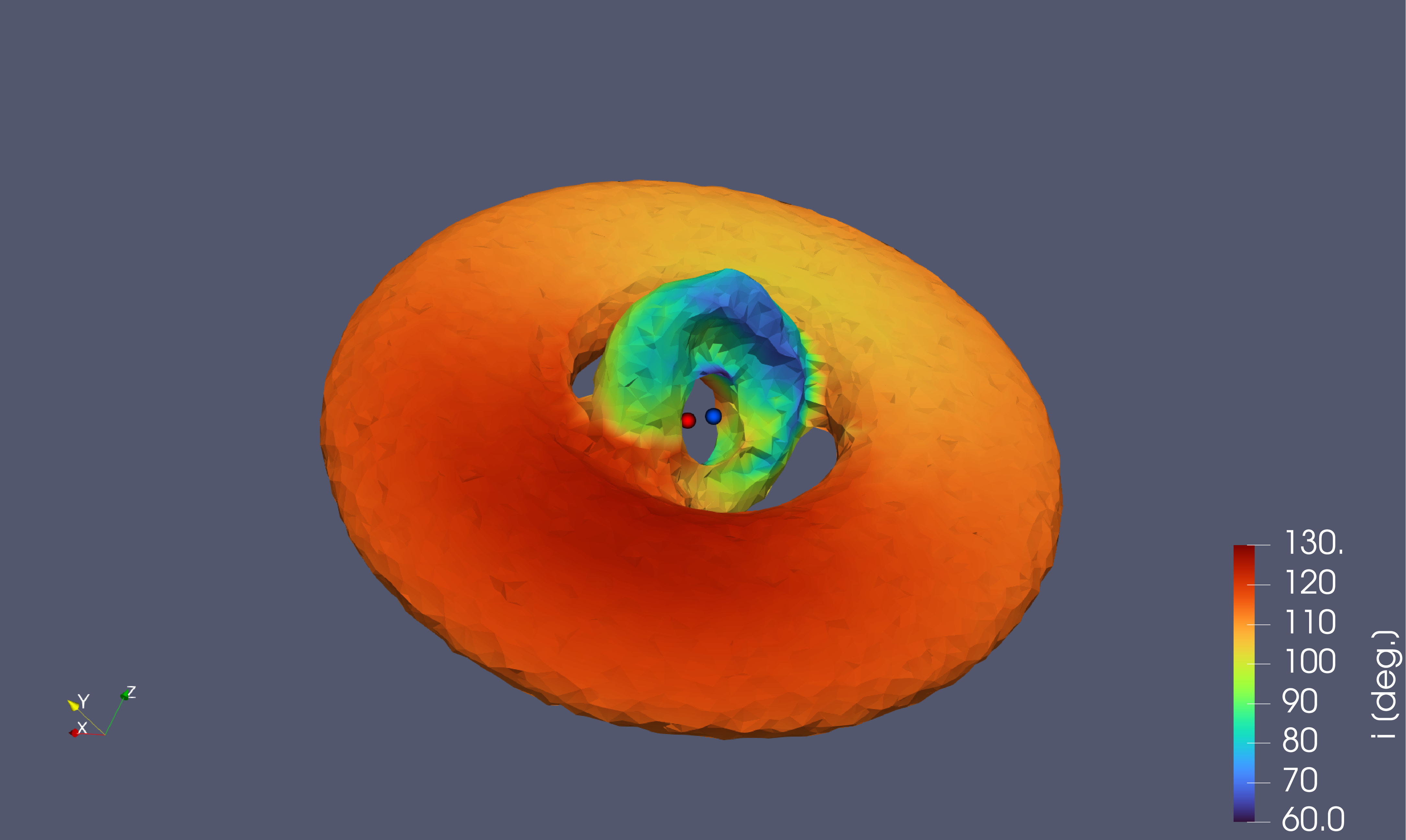}
\includegraphics[width=1\columnwidth]{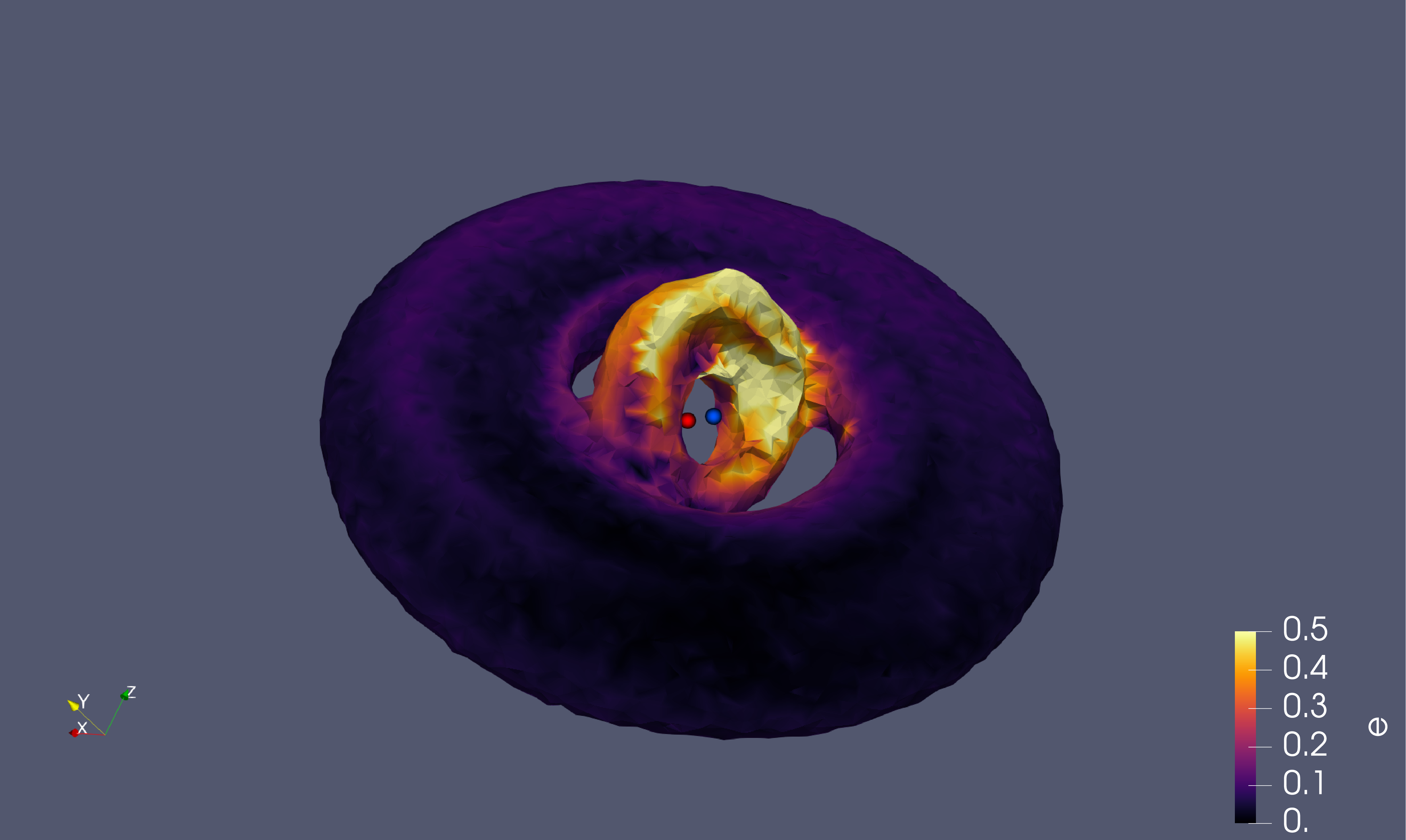}
\includegraphics[width=1\columnwidth]{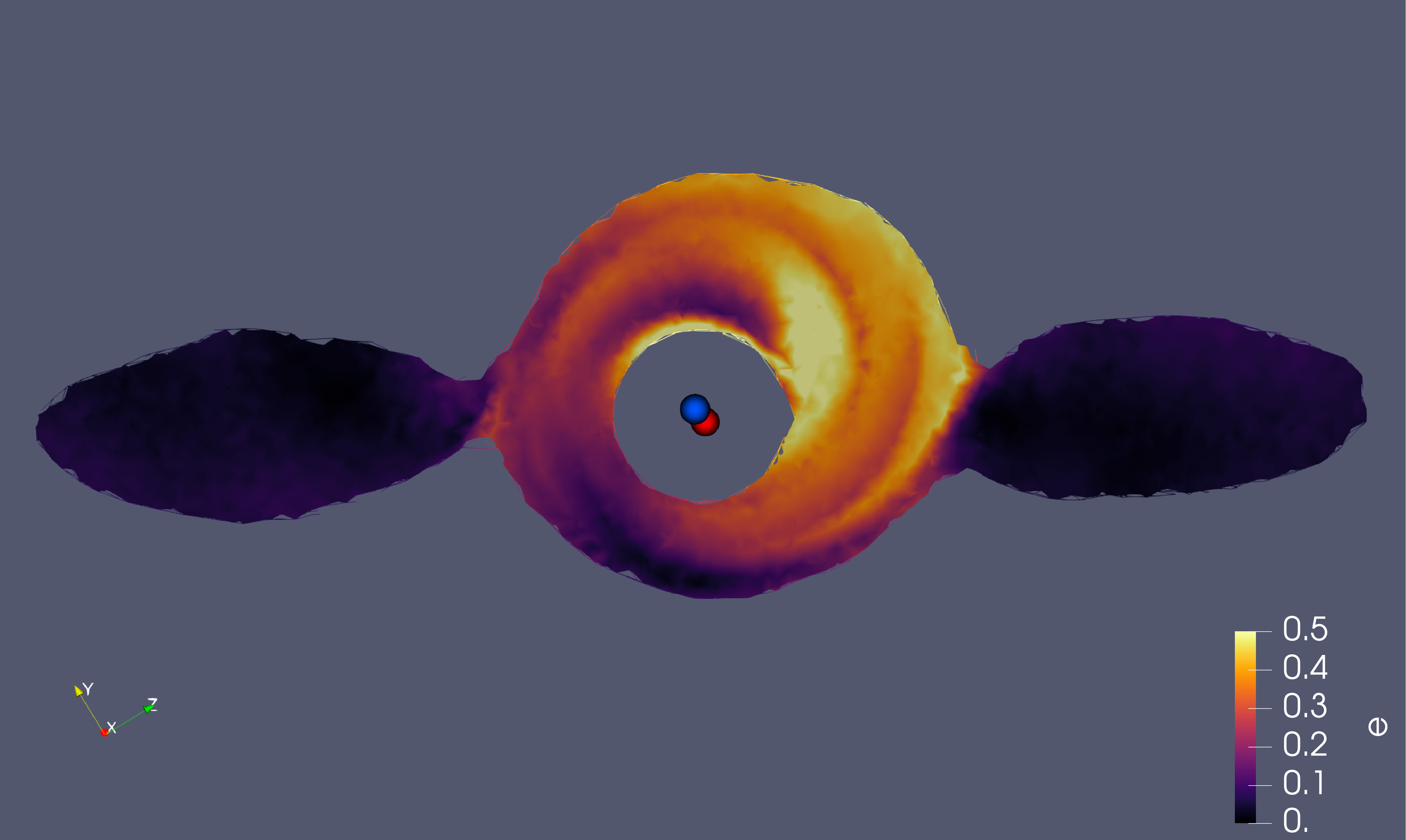}
\end{center}
\caption{A 3D rendering of the circumbinary disc with $i_0 = 45^\circ$ at a time $t = 800\, \rm P_{orb}$ when the disc is breaking. The top panel shows the disc tilt with respect to the binary orbital plane. The middle panel shows the instantaneous disc eccentricity, and the bottom panel shows a cross section of the instantaneous disc eccentricity viewed face-on to the inner disc. The primary star is shown by the blue sphere, while the secondary star is denoted by the red sphere. The size of the spheres corresponds to the sink accretion radii.}
\label{fig::3d_i45}
\end{figure}

% \begin{figure} 
% \begin{center}
% \includegraphics[width=\columnwidth]{figures/i135_tilt.pdf}
% \includegraphics[width=\columnwidth]{figures/i135_ecc.pdf}
% \includegraphics[width=\columnwidth]{figures/i135_ecc_slice.pdf}
% \end{center}
% \caption{Same as Fig.~\ref{fig::3d_i45}, but for the circumbinary disc with $i_0 = 135^\circ$ at a time $t = 1600\, \rm P_{orb}$.}
% \label{fig::3d_i135}
% \end{figure}

We examine the evolution of the disc surface density and eccentricity during the disc alignment process. Figure~\ref{fig::ecc} shows the disc surface density (upper subpanel) and eccentricity (lower subpanel) as a function of time for each model.  The large figure denotes the coplanar case, showing that the the surface density is smooth in radius, with the disc viscously expanding outwards. There is a small eccentricity growth that occurs within the discs due to the binary exciting eccentric modes \cite[e.g.,][]{Lubow2000,Munoz2016,Munoz2019}.
The inclined discs models undergoing coplanar alignment, i.e., $i_0 \leq 30^\circ$, show similar results to the coplanar disc case. Next, we analyze the discs that are undergoing polar alignment. For $i=45^\circ$, the surface density profile reveals disc breaking at $r\sim 2a$. %as the material viscously accretes inward. 
Due to the initial disc tilt being close to the critical tilt for circulation or libration, the disc becomes strongly warped, forcing the disc to break. As the disc breaks, the inner disc becomes very eccentric, reaching an azimuthally-averaged eccentricity of $e\sim 0.2$. Eventually, the break dissipates as the inner and outer discs both align polar. At this point, the disc eccentricity begins to decrease. For $i_0 = 60^\circ$, there is strong warping that occurs as the disc aligns to a polar state, but not strong enough to cause disc breaking. During the periods of strong warping, there is eccentricity growth that occurs but eventually dampens. For $i_0=75^\circ$, there is very little disc warping and eccentricity growth since the initial disc tilt is close to polar. Likewise, for $i_0 = 90^\circ$, the initially polar-aligned disc does not exhibit any strong disc warping or eccentricity growth. 

Next, we examine the discs that are initially retrograde. Like the prograde models, as the initial disc tilt begins closer to the critical tilt, stronger disc warping and larger eccentricity growth occurs. This can be seen for models $i=105^\circ$ and $120^\circ$ in Fig.~\ref{fig::ecc}, where the latter shows stronger warping and eccentricity growth but not enough to cause the disc to break. For $i_0 = 135^\circ$ (close to the critical tilt), the disc breaks, similar to $i_0=45^\circ$, but the time for the disc to break occurs much later. Once the disc is broken, the inner disc becomes very eccentric. For $i_0 = 150^\circ$, the disc undergoes warping {without breaking while its eccentricity increases. Lastly, for the models $i_0 = 165^\circ$ and $180^\circ$, minimal eccentricity growth is present (especially for the $i_0 = 180^\circ$ model). From the above analysis, we can see that the disc structure can vary depending on the initial tilt of the circumbinary disc. 
We investigate how the disc structure impacts the accretion rate onto the binary in Section~\ref{sec::acc_rate}.%The evolution of the disc structure will impact the accretion onto the binary, which we show in Section~\ref{sec::acc_rate}.

We now focus on the disc structure for the two models that showed disc breaking, $i_0 = 45^\circ$ and $135^\circ$. Fig.~\ref{fig::splash_i45} shows the circumbinary disc structure for $i_0 = 45^\circ$ at times ranging from $t = 0\, \rm P_{orb}$ to $t = 3000\, \rm P_{orb}$. At early times, $t < 800\, \rm P_{orb}$, the disc precesses about the eccentricity vector of the binary. At $t = 800\, \rm P_{orb}$, the disc begins to break, where the inner disc evolves quickly to a polar configuration, while the outer disc is still misaligned. The inner disc becomes highly eccentric and spreads outwards with time, interacting with the outer disc. %As time progress, the radial extent of the inner disc increases, as the outer disc interacts with the inner disc. 
Eventually, the break dissipates, and the resulting structure is a coherent polar-aligned disc (see $t = 3000\, \rm P_{orb}$). Figure~\ref{fig::splash_i135} shows the circumbinary disc structure for $i_0 = 135^\circ$ at times ranging from $t = 0\, \rm P_{orb}$ to $t = 3000\, \rm P_{orb}$. At $t < 1300\, \rm P_{orb}$, the disc precesses about the eccentricity vector of the binary with marginal warping. At $t = 1400\, \rm P_{orb}$, the disc begins to break, where the inner disc evolves quickly to a polar configuration, while the outer disc is still misaligned. The timescale for the breaking is much longer than the $i_0 = 45^\circ$ model. The inner disc also becomes highly eccentric. As the break dissipates and the whole disc evolves to a coherent polar-aligned disc, the eccentricity begins to decrease.% as the disc circularizes. 

\begin{figure*} 
\begin{center}
    \includegraphics[width=1\columnwidth]{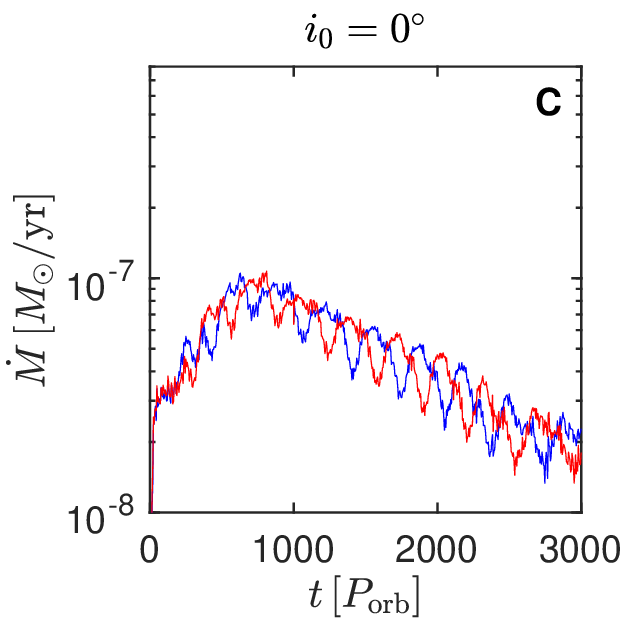}
\end{center}
\includegraphics[width=0.5\columnwidth]{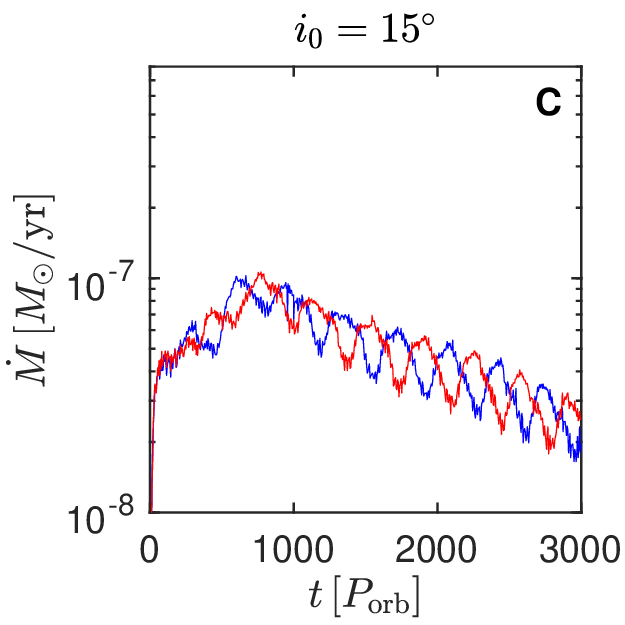}
\includegraphics[width=0.5\columnwidth]{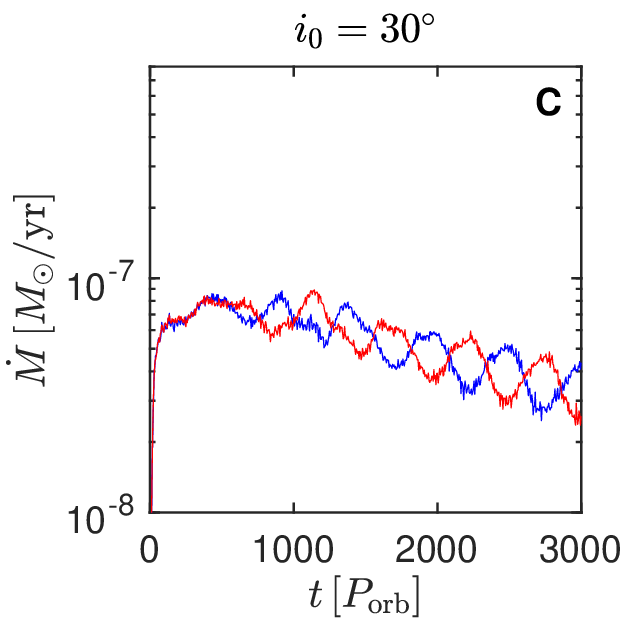}
\includegraphics[width=0.5\columnwidth]{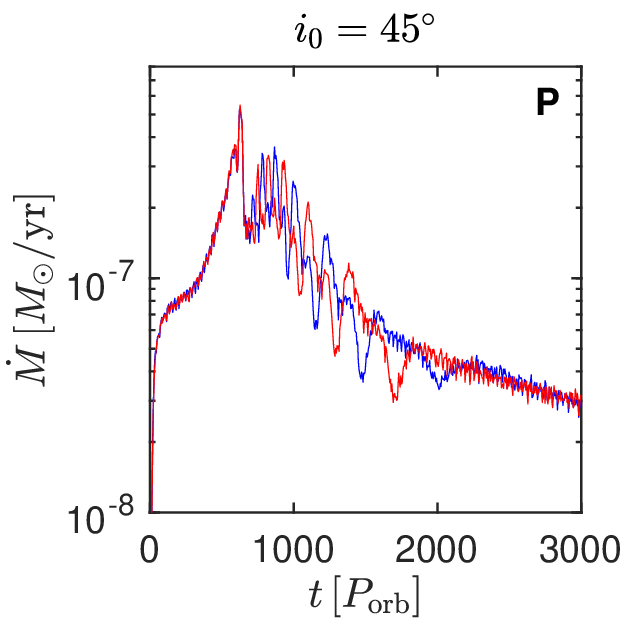}
\includegraphics[width=0.5\columnwidth]{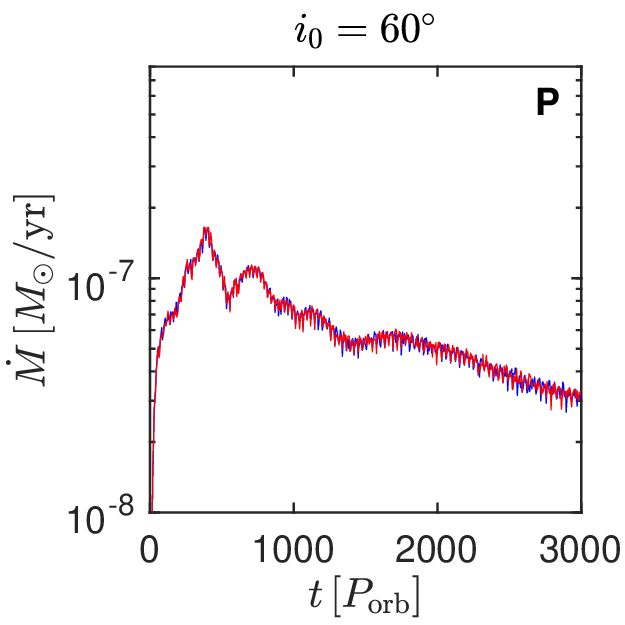}
\includegraphics[width=0.5\columnwidth]{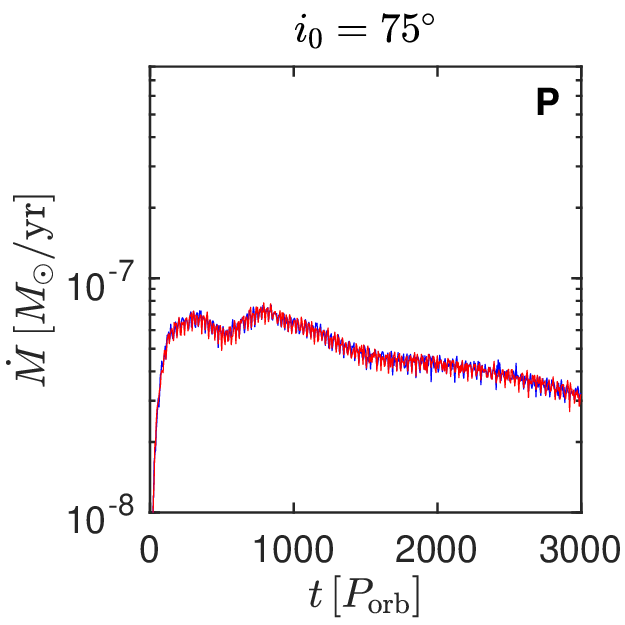}
\includegraphics[width=0.5\columnwidth]{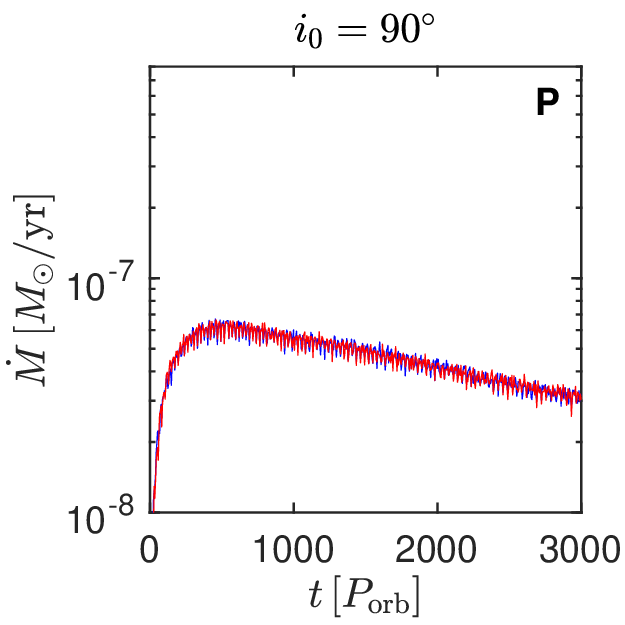}
\includegraphics[width=0.5\columnwidth]{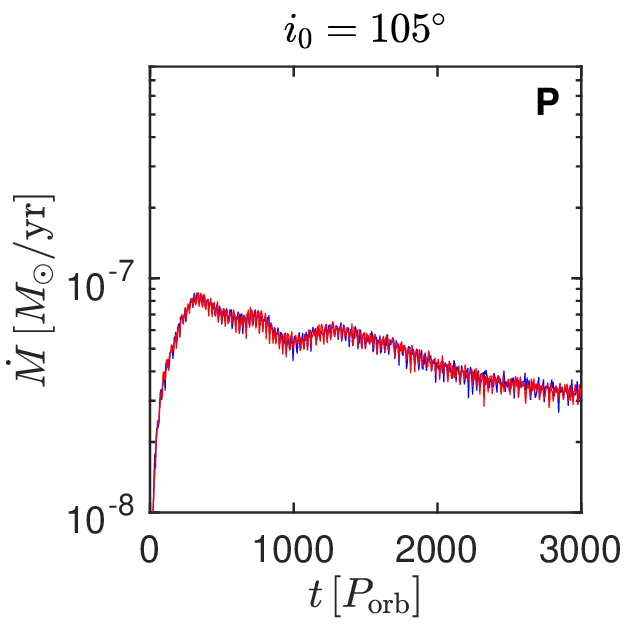}
\includegraphics[width=0.5\columnwidth]{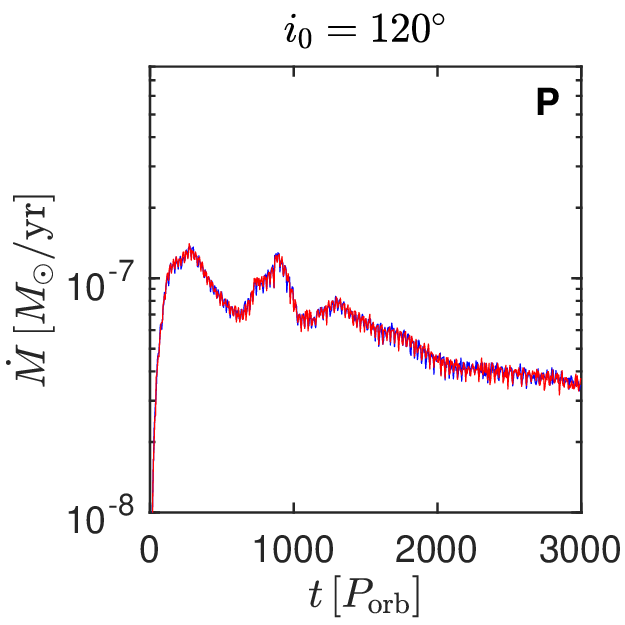}
\includegraphics[width=0.5\columnwidth]{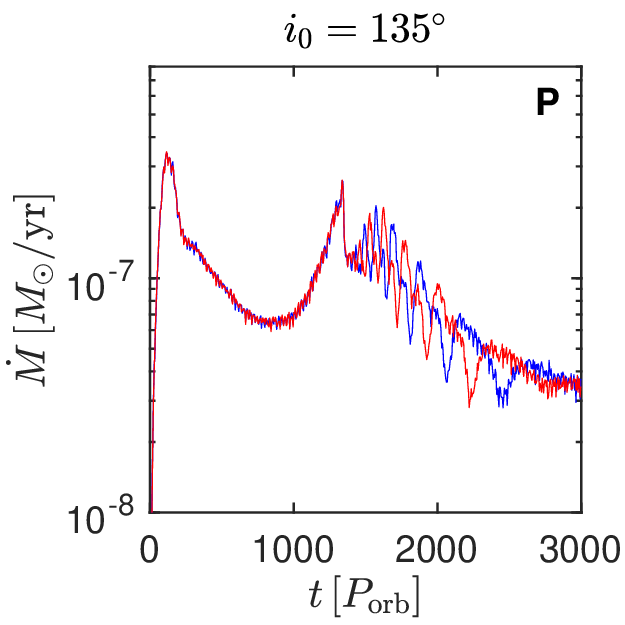}
\includegraphics[width=0.5\columnwidth]{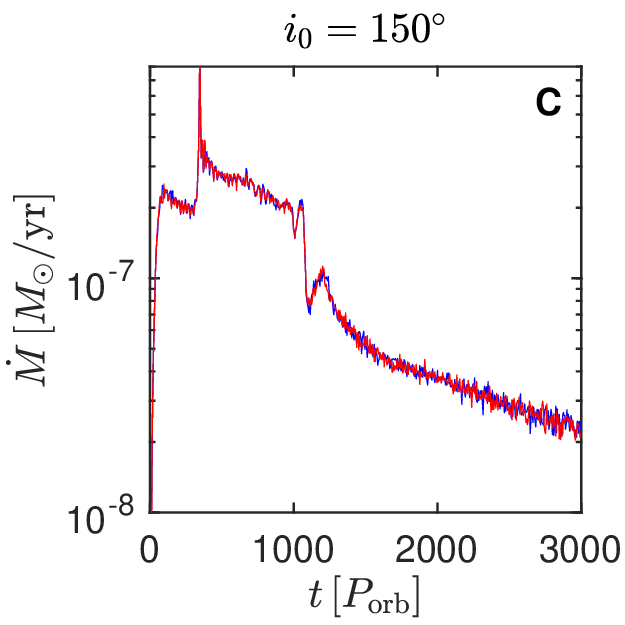}
\includegraphics[width=0.5\columnwidth]{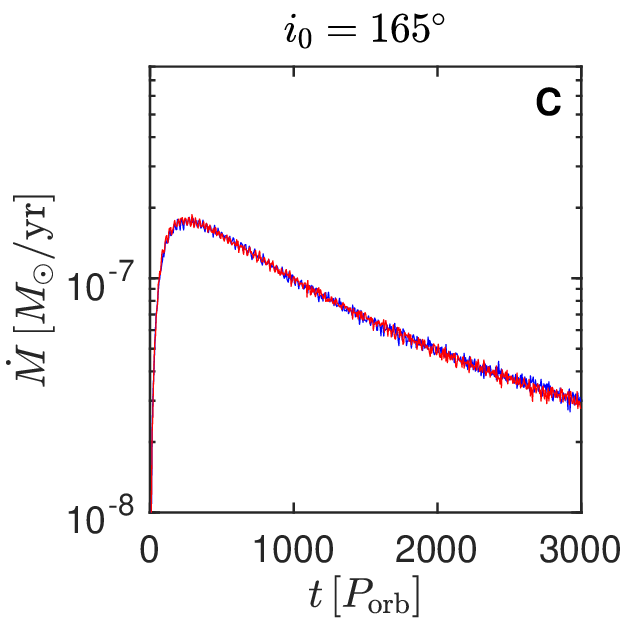}
\includegraphics[width=0.5\columnwidth]{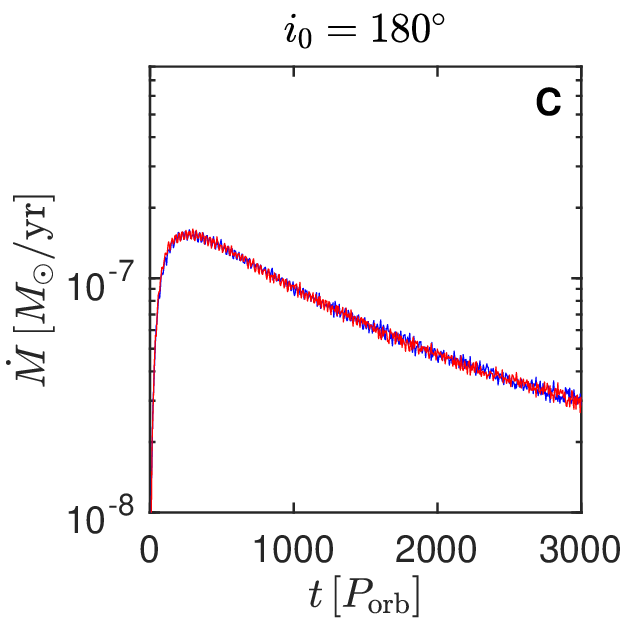}
\caption{The accretion rate, $\dot{M}$, as a function of time in units of binary orbital period, $P_{\rm orb}$, for different initial misalignment of the circumbinary disc (given by the titles). The blue curves represent the primary accretion, while the red curves denote the secondary accretion. The letters "C" or "P" denote whether the disc is undergoing coplanar or polar alignment, respectively. 
% \af{if it is not too much work maybe it would be nice to have a small zoom-in inset in the panels to show a couple of alternating oscillations.} 
}
\label{fig::acc_rate}
\end{figure*}

\begin{figure*} 
\centering
\includegraphics[width=2\columnwidth]{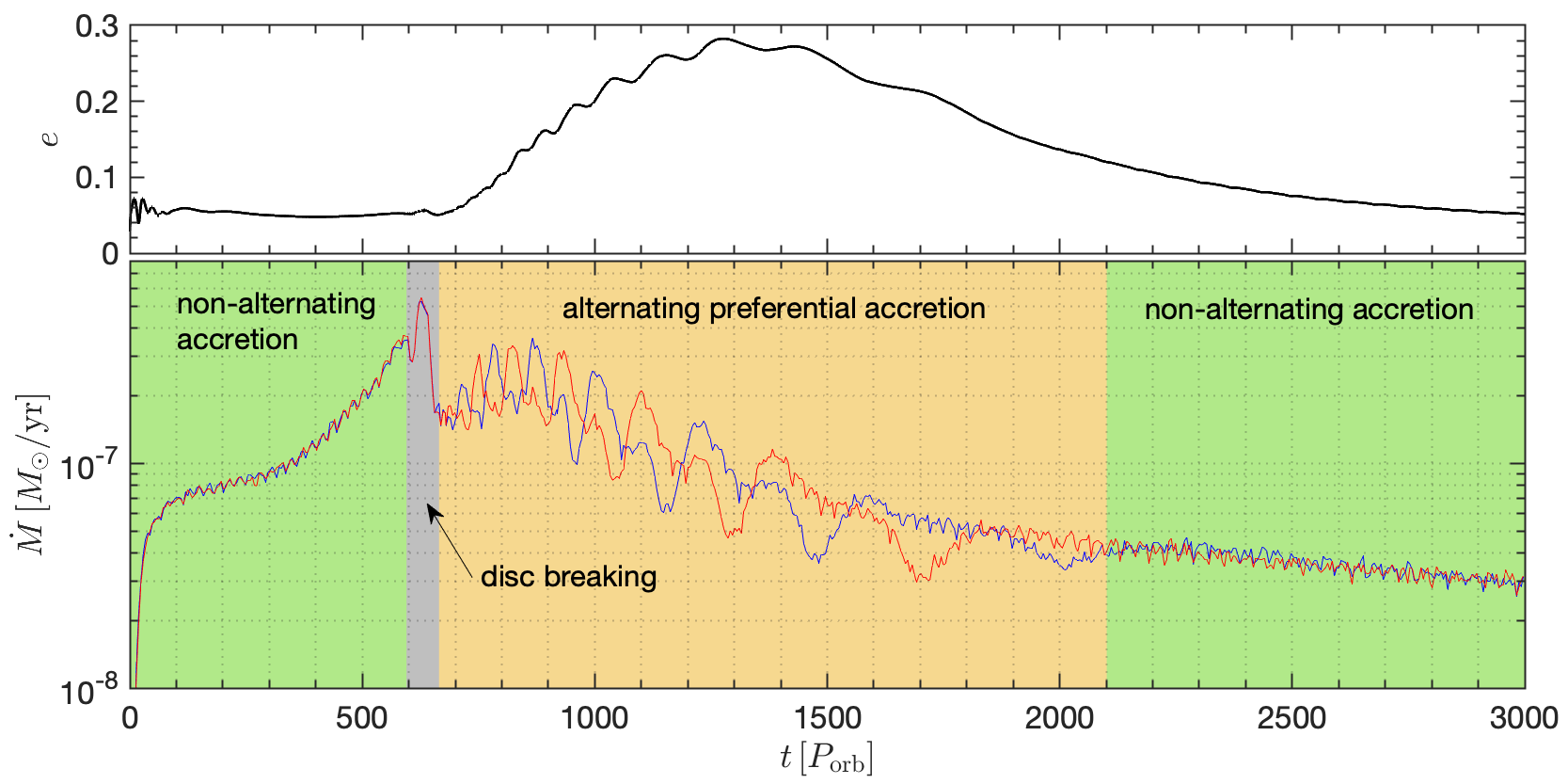}
\caption{Upper panel: the density-weighted average disc eccentricity as a function of time in binary orbital periods, $\rm P_{orb}$. Lower panel: the accretion rate, $\dot{M}$, onto a binary from a circumbinary disc initially misaligned by $i_0 = 45^\circ$ with respect to the binary orbital plane as function of time in binary orbital periods, $\rm P_{orb}$. The green shaded regions denoted when the binary is undergoing non-alternating accretion, where the primary accretion (blue) and secondary accretion (red) are similar. The grey shaded regions represents the time when the disc breaks as it aligns in a polar configuration. The yellow shaded regions shows the time when the binary is undergoing alternating preferential accretion as a result of the disc breaking.}
\label{fig::i45_acc}
\end{figure*}

\begin{figure*} 
\centering
\includegraphics[width=2\columnwidth]{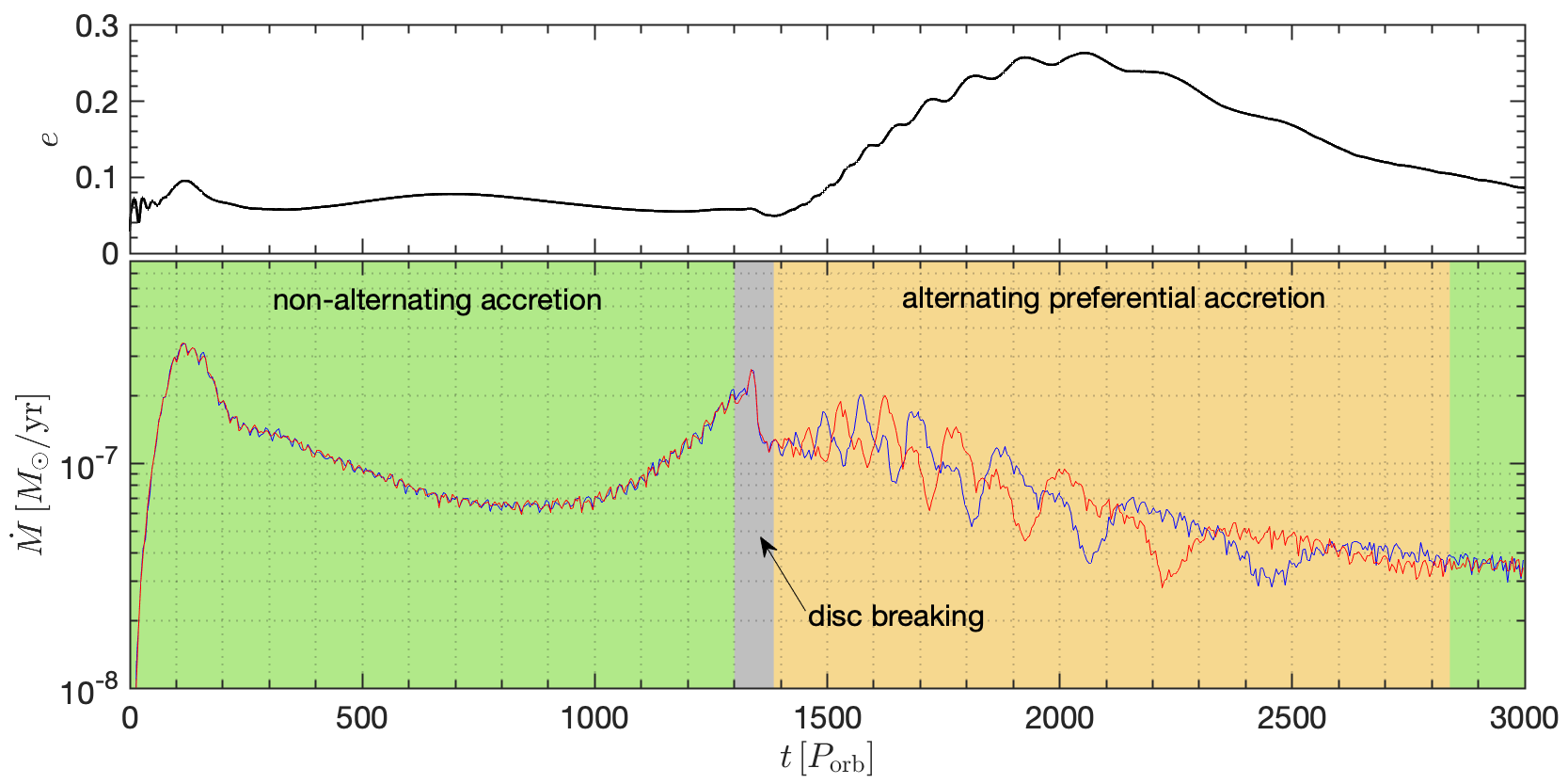}
\caption{Same as Fig.~\ref{fig::i45_acc}, but for a initial circumbinary disc tilt of $i_0 = 135^\circ$.}
\label{fig::i135_acc}
\end{figure*}

We now analyse the disc break evolution for $i_0=45^\circ$. We note that the case for $i_0=135^\circ$ is qualitatively similar to the $i_0=45^\circ$ model. Figure~\ref{fig::i45_disc_params} shows the evolution of the surface density, tilt, longitude of the ascending node, and eccentricity as a function of radius at times $t = 800\, \rm P_{orb}$ (black), $900\, \rm P_{orb}$ (blue), $1000\, \rm P_{orb}$ (teal), $1100\, \rm P_{orb}$ (green), $1200\, \rm P_{orb}$ (yellow), and $1300\, \rm P_{orb}$ (red). At $t = 800\, \rm P_{orb}$, the break in the disc occurs at $\sim 4a$, and propagates outward with time. The inner disc aligns quickly to a polar configuration with the binary orbital plane while the outer disc is still misaligned. Similarly, the longitude of the ascending node for the inner disc is aligned to the azimuthal angle of the eccentricity vector of the binary ($\phi \sim 90^\circ$), while the outer disc is misaligned. Eventually, as the break propagates outwards, the entire disc's tilt and longitude of ascending node will be consistent with polar alignment. The inner disc is always more eccentric than the outer disc but begins to circularize with time (refer back to Fig.~\ref{fig::ecc}).

Figure~\ref{fig::3d_i45} presents a comprehensive 3D representation of the broken disc configuration for $i_0 = 45^\circ$ during the time $t = 800, \rm P_{orb}$.  The upper panel shows the tilt of the disc with respect to the binary orbital plane, the middle panel illustrates the variation of the instantaneous disc eccentricity, and the bottom panel provides a cross-sectional view of the instantaneous eccentricity distribution, with the observer viewing the inner disc face-on. %viewed face-on to the inner disc. 
%Notably, at this particular juncture, 
We note that the inner disc is aligning polar more quickly than the outer disc. A discernible pattern emerges within the inner disc, characterized by substantial growth in eccentricity, with eccentricity values diminishing as a function of radial distance. Remarkably, the inner disc attains an instantaneous eccentricity levels reaching approximately $\sim 0.5$.
% Figure~\ref{fig::3d_i135} shows the same 3D-rendering but for $i_0 = 135^\circ$ at a time $t = 1600\, \rm P_{orb}$.  The inner disc is aligning to a  polar state on a faster timescale than the outer disc. The disc eccentricity grows at the disc is breaking due to the binary torque. 

\subsection{Accretion rate}
\label{sec::acc_rate}

We examine the accretion rates onto the primary and secondary stars for each SPH simulation. Figure~\ref{fig::acc_rate} shows a summary of the accretion rate, $\dot{M}$, onto the primary (blue) and secondary (red) stars as a function of time in binary orbital periods, $\rm P_{orb}$.  The accretion onto the primary and secondary stars exhibit short-term variability with timescales similar to the binary orbital period, as expected from previous simulations  of coplanar eccentric circumbinary discs \citep{Farris2014,Munoz2020a,Duffell2020,Franchini2023,Lai2023}. For the disc undergoing prograde coplanar alignment, the accretion rates onto the primary and secondary exhibit antiphase behavior. This means that when the accretion rate onto the primary is at its peak, the accretion rate onto the secondary is at its lowest. This alternating preferential accretion is present even if the circumbinary disc is tilted and aligning coplanar to the binary orbital plane.  For the initial coplanar disc orientation ($i_0 = 0^\circ$), the modulation of the alternating preferential accretion is related to the apsidal precession rate of the disc. In the limit of a coplanar pressure-less particle disc, the secular apsidal precession rate around an eccentric binary is
\begin{equation}
    \dot{\omega}_d \simeq \frac{3\Omega_{\rm b}}{4}\frac{q_{\rm b}}{(1+q_{\rm b})^2}\bigg( 1+\frac{3}{2}e_{\rm b}^2 \bigg) \bigg( \frac{a_{\rm b}}{r}\bigg)^{7/2}
\end{equation}
\citep{Munoz2016}, where $q_{\rm b} = M_2/M_1$. At $r\sim 3a$, the precession period is $\sim 200\, \rm P_{orb}$. Similar oscillations with a consistent time-scale were observed in prior works \citep{Dunhill2015,Munoz2016,Munoz2019,Lai2023}.
%For an inclined disc, the torque exerted onto the disc from the binary is weaker, which causes the modulation of the alternating preferential accretion to be longer.

Next, we analyze the accretion rates for the disc undergoing polar alignment. The further the initial tilt of the disc is from polar, the more eccentric the disc becomes, which in turn causes oscillations in the accretion rates. If the initial tilt is close to the critical tilt for polar alignment, the large eccentricity growth causes significant fluctuations in the accretion rates, as seen in models with $i_0 = 45^\circ$ and $i_0=135^\circ$. These two models exhibit alternating preferential accretion for a period of time in a polar-aligning disc. The remaining polar-aligning models show no evidence of alternating preferential accretion; that is, the primary accretion is similar to the secondary accretion. For an initially polar disc, there are no oscillations in the accretion rates since there is little eccentricity growth present in this type of initial configuration \cite[e.g.,][]{Smallwood2022}. Lastly, we examine the accretion rates for discs undergoing retrograde coplanar alignment. These models do not show any alternating preferential accretion. However, for $i_0 = 150^\circ$, there are oscillations in the accretion rate due to strong disc warping. From the summary of the accretion rates from our suite of SPH simulations, it is evident that the accretion rate evolution can be affected by the initial tilt of the circumbinary disc.

We further analyze the accretion rates for the two models that exhibit temporary alternating preferential accretion for a polar-aligning circumbinary disc, namely $i_0 = 45^\circ$ and $i_0 = 135^\circ$. Figure~\ref{fig::i45_acc} illustrates the accretion rate, $\dot{M}$, onto the binary from a circumbinary disc initially misaligned by $i_0 = 45^\circ$, plotted as a function of time in binary orbital periods, $\rm P_{orb}$. Prior to the disc breaking, the accretion rate displays non-alternating behavior, with primary accretion similar to secondary accretion, as indicated by the green shaded region. Around $t = 800, \rm P_{orb}$, the disc breaks due to the strong warping induced by the binary during polar alignment, shown by the grey shaded region. As the disc breaks, the inner disc becomes eccentric (shown in the upper panel, and also refer to Figs.~\ref{fig::ecc} and~\ref{fig::3d_i45}). Consequently, the accretion rate onto the binary shifts from non-alternating accretion to alternating preferential accretion, highlighted by the yellow shaded region. It is noteworthy that during this period of alternating preferential accretion, the inner disc is nearly in a polar configuration. As the disc eccentricity decreases again, the accretion rate transitions back from alternating preferential accretion to non-alternating accretion. The $i_0 = 135^\circ$ model shows similar behaviour, as shown in Fig.~\ref{fig::i135_acc}. The main difference is that the disc breaking and period of alternating preferential accretion occurs at a later time. The accretion rate then switches back to non-alternating accretion as the disc eccentricity decreases.

% \begin{figure} 
% \centering
% \includegraphics[width=\columnwidth]{figures/binary_a.eps}
% \caption{XX}
% \label{fig::binary_a}
% \end{figure}

% \begin{figure} 
% \centering
% \includegraphics[width=\columnwidth]{figures/binary_e.eps}
% \caption{XX}
% \label{fig::binary_e}
% \end{figure}

% \section{Binary evolution}
% \label{sec::binary}

\section{Discussion}
\label{sec::discussion}

\subsection{Disc breaking}
From Fig.~\ref{fig::ecc}, as the initial disc tilt is near the critical tilt for polar alignment, the disc breaks, causing a period of alternating preferential accretion. The break is roughly located at $4a$, which is equal to the initial inner radius of the disc in our simulations. However, disc material will flow inward, extending closer to the binary. The inner disc radius of a circumbinary disc is determined by the balance between the viscous torque exerted by the circumbinary disc and the tidal/resonant torque of the binary \citep{Artymowicz1994}. Under conditions of moderate orbital eccentricity, the typical inner edge resides near the 1:3 outer Lindblad resonance radius, approximately estimated to be $\sim 2a$. However, if the circumbinary disc is tilted relative to the binary orbital plane, the resonant torque weakens, resulting in a reduction of the inner disc edge \citep{Franchini2019b}. When the inner radius decreases, the accumulation of material in the inner regions increases, potentially leading to a reduction in the precession timescale compared to the radial communication timescale, ultimately resulting in disc breaking.

The mechanism by which a warp is communicated through the disc depends on the \cite{Shakura1973} disc viscosity $\alpha_{\rm SS}$, and aspect ratio $H/r$. In regimes where $\alpha_{\rm SS} < H/r$, which is tipically relevant for protoplanetary discs, the warp propagates as a bending wave \citep{Papaloizou1995,Lubow2000}. When $\alpha_{\rm SS} > H/r$, the warp propagation occurs in the diffusive regime which is applicable to thin, fully ionized discs. When warps become very large, the disc may be unable to communicate across the warp effectively. The warp may become unstable and disc breaking can occur, splitting the disc into separate rings. For our disc-breaking simulations, strong warping and a high nodal precession timescale force the disc to break. However, predicting the location of the breaking radius is still unclear. The disc breaking phenomenon has been studied analytically \citep{Dovgan2018,Raj2021} and numerically in the context of single and binary black holes \citep{Nixon2012,Nixon2013,Nealon2015}. Disc breaking may also be an explanation for observed disc geometries and accretion kinematics \citep{Casassus2015,Facchini2018,Zhu2019,Kraus2020,Nealon2022}. Disc breaking is more likely for discs with smaller inner cavities, cooler temperatures, and steeper power-law profiles \cite[e.g.,][]{Rabago2023}. Analytical criterion for disc breaking has been shown by \cite{Dovgan2018} and \citep{Deng2022}.

Viscosity can play a significant role in the conditions for disc breaking to occur \cite[e.g.,][]{Drewes2021,Deng2022,Young2023,Rabago2023}. \cite{Young2023} modelled warping and disc breaking around binary systems, finding the breaking can occur for a misalignment of $\lesssim 40^\circ$ for moderate binary eccentricities in the bending-wave regime. For high binary eccentricities, $\sim 0.8$, they found the disc is more susceptible to breaking for lower misalignment angles. \cite{Deng2022} found that the disc broke at much smaller misalignment angle ($\sim 14^\circ$), which can be attributed to their simulated disc viscosity, as they modeled an "inviscid" disc. In the current study, we model slightly viscous discs, similar to \cite{Young2023}. However, if we model lower viscosities, approaching the inviscid regime, it may be possible that the strongly warped models in our suite of SPH simulations that did not break, will break -- leading to temporary periods of alternating preferential accretion.

\subsection{Resolution}
The smoothed particle hydrodynamics (SPH) method is predominantly utilized in simulations of misaligned circumbinary discs. This method incorporates explicit artificial viscosity, which, particularly at lower resolutions, may lead to a significant effective viscosity. Therefore, sufficient resolution is needed to accurately model a wave-like disc \cite[e.g.,][]{Drewes2021}. \cite{Drewes2021} illustrated that when employing sufficiently high resolution, the numerical viscosity remains low enough to attain reasonable agreement between the outcomes of {\sc phantom} SPH simulations in the linear regime and the linearized wave-like warp evolution equations proposed by \cite{Lubow2000}.

Figure~\ref{fig::resolution} depicts the average smoothing length per scale height, $\langle h \rangle /H$, as a function of disc radius, $r$. We present the same times as in Fig.~\ref{fig::i45_disc_params}, corresponding to when the disc is broken. Smaller $\langle h \rangle /H$ means higher resolution, while larger $\langle h \rangle /H$ means lower resolution. The peaks in resolution coincide with the location of the breaking radius. The inner and outer discs are consistently well-resolved. However, the resolution worsens closer to the cavity. At the breaking radius, the resolution is $\langle h \rangle /H < 0.5$, indicating moderate resolution at the breaking radius. The magnitude of the accretion rates analysed in Fig.~\ref{fig::acc_rate} depends on resolution. For example, for lower resolution, $\langle h \rangle /H$ may be above unity which will increase the artificial viscosity and thus artificially increase the accretion rates. This effect can be seen in Fig.~6 in \cite{Smallwood2023}, where the lower resolution simulation has slightly higher accretion rate than the higher resolution simulation. At smaller radii in Fig.~\ref{fig::resolution}, the disc becomes increasingly unresolved, suggesting that the gaseous streams accreting onto the binary components are not well resolved. However, it's important to note that the focus of this work is on the accretion rate pattern and its evolution, rather than on the absolute values of the accretion rates. Therefore our results on preferential accretion alternation are not affected by the low resolution we have in the breaking discs cases.

\begin{figure} 
\centering
\includegraphics[width=\columnwidth]{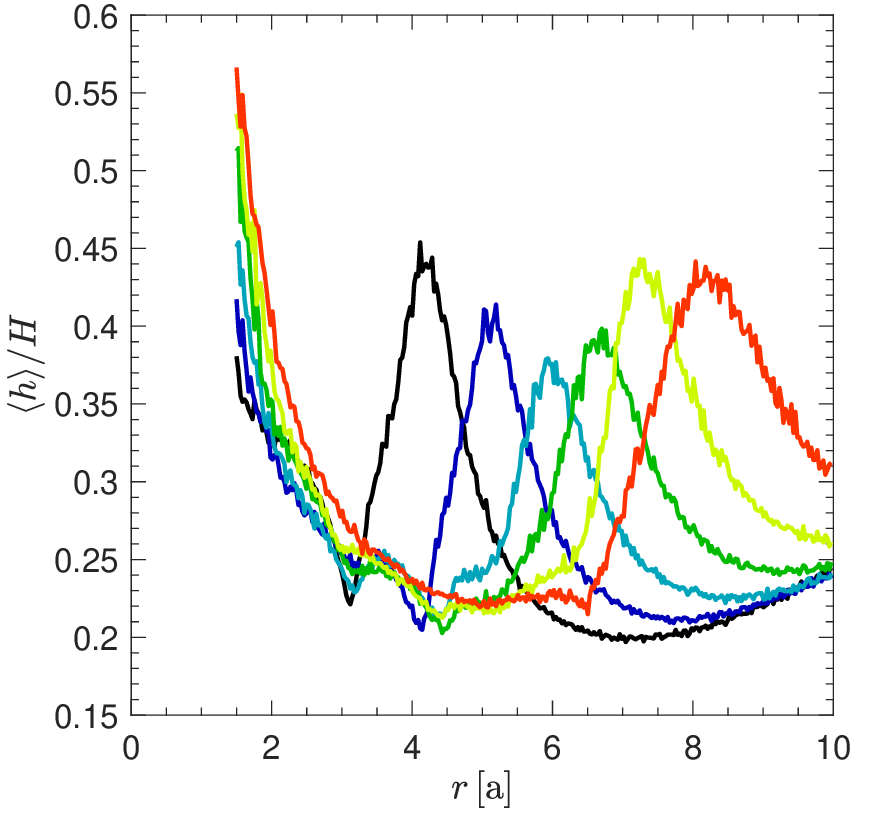}
\caption{ The average smoothing length per scale height, $\langle h \rangle /H$, as a function of disc radius, $r$ for model of $i_{0}=45^{\circ}$. We show selected times beginning with $t = 800\, \rm P_{orb}$ (black), $900\, \rm P_{orb}$ (blue), $1000\, \rm P_{orb}$ (teal), $1100\, \rm P_{orb}$ (green), $1200\, \rm P_{orb}$ (yellow), and $1300\, \rm P_{orb}$ (red). The peak in the resolution is located at the breaking radius.}
\label{fig::resolution}
\end{figure}

\subsection{Observational implications}

%{\bf Accretion observations in binary star systems often focus on the emission lines, continuum emission, or variability that results from the accretion process. }
 Spectroscopy is one of the primary tools used to study accretion in binary systems. The VLT X-shooter spectrograph is an excellent instrument for measuring accretion rates in binary star systems, thanks to its broad wavelength coverage from ultraviolet (UV) to near-infrared (NIR) (300–2500 nm) \citep{Vernet2011}. This range enables the simultaneous observation of key accretion indicators, such as  $\rm H\alpha$, Ca II infrared triplet, and $\rm Br\gamma$ emission lines, which trace ionized gas accreting onto the stars \citep{Hartmann1998}. Additionally, X-shooter’s capability to detect UV excess, particularly in the Balmer continuum, allows for the determination of accretion luminosity, a crucial step in calculating accretion rates \citep{Rigliaco2012}. In binary systems, where the spectra of two stars may overlap, X-shooter’s high spectral resolution enables the decomposition of individual stellar components, allowing for precise measurements of accretion rates for each star in the system \citep{Alcala2014,Manara2016,Manara2017}. This multi-wavelength approach is particularly effective in understanding complex accretion dynamics in young stellar systems. 

 Accretion rates variability can also be probed using photometric monitoring \cite[e.g.,][]{Jensen2007}. Changes in brightness, often in the ultraviolet (UV) or optical bands, are linked to accretion hot spots on the stellar surface, which form where gas from the disc impacts the star. These hot spots increase in brightness as the accretion rate intensifies, causing periodic or irregular variability in the light curves \citep{Cody2010,Kurosawa2013,Venuti2014}. In binary systems, this variability can also be modulated by the orbital motion. For example, the low-mass, pre-main-sequence eclipsing binary 2M1222--57 exhibits active accretion, as evidenced by modulated H$\alpha$ emission \citep{Stassun2022}. While observations indicate a dominant period of $P_{\rm b}$, simulations of a nearly circular, equal-mass ratio binary predict a dominant period of $5P_{\rm b}$. This discrepancy suggests that the unequal mass ratio in 2M1222--57 plays a significant role \cite[see Section 5.3 in][]{Stassun2022}. The dominant variability of $5\ P_{\rm b}$ is associated to the disc inner edge motion, which is where the accretion streams originate from \citep{Munoz2016}.

% on the orbital period for nearly circular, equal-mass ratio binary, \textcolor{red}{the observations showed a dominant period at $P_{\rm b}$, but the simulations of a nearly circular, equal-mass ratio binary predicated that the dominant period is at $5P_{\rm b}$. So the authors suggested that the unequal mass ratio for 2M1222--57 matters. See discussions at the end of Section 5.3 in their paper.} 
% the dominant variability is at $5\ P_{\rm b}$ \citep{Stassun2022}, which is associated to the disc inner edge motion, which is where the streams originate from \citep{Munoz2016}.
%is consistent with accretion streams originating from the circumbinary disc \citep{Munoz2016}.

% \textcolor{red}{This statement appears in the next paragraph: Additionally, based on time-series photometry and high-resolution optical spectroscopy, \cite{Tofflemire2019} found the circumbinary accretion onto the TWA 3A peaks during periastron passages and shows evidence of preferential accretion onto the primary star, TWA 3Aa. }
% \textcolor{red}{duplicated as it appears below: More recently, it was found by \cite{Czekala2021} that the circumbinary disc around TWA 3A is in a nearly coplanar configuration with a tilt $<6^\circ$ with respect to the binary orbital plane.}

 An example of a system that may show evidence of alternating preferential accretion is TWA 3. The TWA 3 system consists of three pre-main sequence stars \cite[$10\pm 3\, \rm Myr$,][]{Bell2015} in a hierarchical configuration. The inner binary, TWA 3A, has an orbital period of $P = 34.8785\pm 0.0009$ days, an eccentricity of $e = 0.628 \pm 0.006$, and a binary mass ratio of $q = 0.841\pm 0.014$. Based on time-series photometry and high-resolution optical spectroscopy, \cite{Tofflemire2019} found the circumbinary accretion onto the TWA 3A peaks during periastron passages and shows evidence of preferential accretion onto the primary star, TWA 3Aa. However, preferential accretion switching back to the secondary is not yet currently observed. Due to the short period of the binary orbit, we may be able to observe the long-term preferential accretion behavior within a reasonable timescale in the future. More recently, it was found by \cite{Czekala2021} that the circumbinary disc around TWA 3A is in a nearly coplanar configuration with a tilt $<6^\circ$ with respect to the binary orbital plane. This is consistent with our hydrodynamical simulations that a lowly inclined circumbinary disc undergoing coplanar alignment will exhibit preferential accretion onto the primary for a period of time.

 High-angular-resolution techniques such as interferometry have become crucial for resolving the structures of accretion discs. Instruments like the Very Large Telescope Interferometer (VLTI) and Atacama Large Millimeter/submillimeter Array (ALMA) can spatially resolve the circumbinary and circumstellar discs \citep{Kraus2017,Kennedy2019,Corporaal2021}. These observations allow for direct imaging the gas dynamics and disc structures, providing insight into how material is funneled toward one or both stars \citep{Alves2019}. For example, \cite{Bohn2022} used a combination of VLTI and ALMA observations to study the misalignments in transitions discs.  Such observations may be applied to understanding binary systems, as they can provide insights into how accretion flows behave in systems where the disc is not co-planar with the binary orbit, like the systems detailed in \cite{Czekala2019} and \cite{Ceppi2024}.

\section{Conclusions}
\label{sec::conclusion}

Our suite of hydrodynamical simulations shed light on the intricate dynamics of material accretion from circumbinary discs in binary star systems, particularly when considering the binary-disc misalignment. We conducted SPH simulations varying initial tilts from $0^\circ$ to $180^\circ$, encompassing coplanar prograde, polar, and coplanar retrograde alignments. Our findings reveal distinctive accretion patterns contingent upon the initial disc tilt. Discs evolving towards prograde coplanar alignment exhibit alternating preferential accretion onto the primary and secondary stars, while those approaching polar alignment may experience disc breaking and subsequent transition to alternating preferential accretion due to strong disc warping. Conversely, discs undergoing retrograde coplanar alignment do not incite alternating preferential accretion. These results demonstrate the pivotal role of initial disc tilt in shaping accretion rate evolution, which can may be used as a diagnostic  discerning binary-disc misalignment.

The results from this work may also be applied to super-massive black hole binaries (SMBHBs) as our simulations are scale free. Since galaxy mergers are expected to channel large amounts of gas toward the center of the merger remnant \citep{Hopkins2006}, gaseous circumbinary discs should naturally occur around SMBHBs. Measuring the accretion rates of SMBHBs effectively \cite[e.g.,][]{DOrazio2024}, and showing possible signs of alternating preferential accretion may shed light on the system architecture. %Circumbinary accretion in the context of SMBHs may be a solution to the "final parsec problem" \citep{Milosavljevic2003a,Milosavljevic2003b}, whereby gaseous accretion can shrink the SMBH separation \cite[e.g.,][]{Nixonetal2011a,Volonteri2020,Volonteri2022}.

%%%%%%%%%%%%%%%%%%%%%%%%%%%%%%%%%%%%%%%%%%%%%%%%%%
\section*{Acknowledgements}
 We thank the anonymous referee for helping to improve the quality of the manuscript. JLS thanks Diego J. Mu\~{n}oz for helpful suggestions that improved the quality of the manuscript. JLS acknowledges funding from the ASIAA Distinguished Postdoctoral Fellowship and the Taiwan Foundation for the Advancement of Outstanding Scholarship. Y.P.L. is supported in part by the Natural Science Foundation of China (grants Nos. 12373070, and 12192223), the Natural Science Foundation of Shanghai (grant No. 23ZR1473700).

%%%%%%%%%%%%%%%%%%%%%%%%%%%%%%%%%%%%%%%%%%%%%%%%%%
\section*{Data Availability}

The data supporting the plots within this article are available on reasonable request to the corresponding author. A public version of the {\sc phantom} and {\sc splash} codes are available at \url{https://github.com/danieljprice/phantom} and \url{http://users.monash.edu.au/~dprice/splash/download.html}, respectively. The 3D renderings made use of {\sc paraview}, which is available at \url{https://www.paraview.org/}.

%%%%%%%%%%%%%%%%%%%% REFERENCES %%%%%%%%%%%%%%%%%%

% The best way to enter references is to use BibTeX:

\bibliographystyle{mnras}
\bibliography{ref.bib} % if your bibtex file is called example.bib

% Alternatively you could enter them by hand, like this:
% This method is tedious and prone to error if you have lots of references
%\begin{thebibliography}{99}
%\bibitem[\protect\citeauthoryear{Author}{2012}]{Author2012}
%Author A.~N., 2013, Journal of Improbable Astronomy, 1, 1
%\bibitem[\protect\citeauthoryear{Others}{2013}]{Others2013}
%Others S., 2012, Journal of Interesting Stuff, 17, 198
%\end{thebibliography}

%%%%%%%%%%%%%%%%%%%%%%%%%%%%%%%%%%%%%%%%%%%%%%%%%%

%%%%%%%%%%%%%%%%% APPENDICES %%%%%%%%%%%%%%%%%%%%%

% \appendix

% \section{Some extra material}

% If you want to present additional material which would interrupt the flow of the main paper,
% it can be placed in an Appendix which appears after the list of references.

%%%%%%%%%%%%%%%%%%%%%%%%%%%%%%%%%%%%%%%%%%%%%%%%%%

% Don't change these lines
\bsp	% typesetting comment
\label{lastpage}
\end{document}